\documentclass[%
 reprint,
nofootinbib,
 amsmath,amssymb,
 aps,
prd,
]{revtex4-2}

\usepackage{graphicx}
\usepackage{dcolumn}
\usepackage{bm}
\usepackage{siunitx}

\usepackage{makecell}
\usepackage{tabularx}
\usepackage{hyperref}
\usepackage{aas_macros}

\newcommand{\msun}{\,\mathrm{M}_\odot}

\begin{document}

\preprint{}

\title{Long-time 3D supernova simulations of non-rotating progenitors with magnetic fields}

\author{Bailey Sykes}
 \email{bailey.sykes@monash.edu}
\author{Bernhard M\"{u}ller}%
 \email{bernhard.mueller@monash.edu}
\affiliation{%
 School of Physics and Astronomy, Monash University, Clayton, VIC, 3010, Australia
}%

\date{\today}

\begin{abstract}
We perform five 3D magnetohydrodynamic (MHD) core-collapse supernova simulations for non-rotating progenitors between $9.5 \,\mathrm{M}_\odot$ and $24 \,\mathrm{M}_\odot$.  Four of the five models produce explosions while one fails. The exploding models are extended to between $0.9 \, \mathrm{s}$ and $1.6 \, \mathrm{s}$ post-bounce to study a possible impact of magnetic fields on explosion and remnant properties. Diagnostic explosion energies grow at a similar pace as in previous non-magnetic models. They reach between $0.11 \, \mathrm{foe}$ and $0.61 \, \mathrm{foe}$, but are still growing by the end of the simulations. Neutron star kicks reach no more than $300\,\mathrm{km}\,\mathrm{s}^{-1}$, and although these are also still growing, they are unlikely to be in conflict with observed pulsar velocities.  Extrapolated neutron star spin periods are between $45\,\mathrm{ms}$ and $1.8\,\mathrm{s}$, consistent with observed birth spin rates. Magnetic torques only contribute about 10\% to the spin-up of the neutron star. The inclusion of magnetic fields does not provide a mechanism for spin-kick alignment in our simulations. Surface dipole fields are in the range of  $10^{12}$-$10^{13}\,\mathrm{G}$, much smaller than the root-mean-square field strength. Different from previous simulations, magnetic fields in the gain region only reach at most $\mathcal{O}(1\%)$ of kinetic equipartition, likely because relatively early shock revival cuts off accretion as a power source for field amplification, which appears to be driven primarily by shear flows at the bottom of the gain region.

\end{abstract}

\maketitle

\section{Introduction}
In recent years, 3D supernova simulations have made enormous progress in understanding the collapse and supernova explosion of massive stars ($M \gtrsim 8 \msun$). Models of neutrino-driven explosions have matured to the degree that they can obtain explosion and remnant properties in rough agreement with transient and compact remnant observations, and are also becoming useful for studying the systematics
of these explosion and remnant properties with progenitor
mass \citep[e.g.,][]{Mueller_et_al:2019,Bollig_et_al:2021,Burrows_Wang_Vartanyan:2024,Janka_Kresse:2024}.

One major construction site that remains in the modern paradigm of the core-collapse supernova (CCSN) explosion mechanism
is the role of magnetic fields \citep{mueller_24}.  
Thus, while the systematics of neutron star birth masses, kicks, and spins are becoming increasingly transparent thanks to modern 3D simulations, the magnetic fields of young neutron stars as another key observable remain a much bigger enigma. The magnetic field of an astrophysical NS can be constrained by observation of its spin period, $P$, and its spin-down rate, $\dot{P}$ \citep{Igoshev_Popov_Hollerbach:2021, Mereghetti_Pons_Melatos:2015}. These estimates indicate a broad spectrum of magnetic field strengths, ranging from $10^{8} \, \mathrm{G}$ for fast radio pulsars, up to $10^{15} \, \mathrm{G}$ for highly magnetized magnetars \citep{Reisenegger:2003, Turolla_Zane_Watts:2015,Kaspi_Beloborodov:2017}. The origin of these fields remains uncertain. Two scenarios, a fossil-field
origin relying on flux conservation during collapse \citep{Ferrario_Wickramasinghe:2006}, and a dynamo origin
\citep{duncan_92,thompson_93}, have been proposed. Although arguments can be made for and against these scenarios based on population statistics \citep{Ferrario_Wickramasinghe:2008,Makarenko_Igoshev_Kholtygin:2021}, better simulations of magnetic field evolution during CCSNe are critical for unraveling the origin and population distribution of neutron star magnetic fields. Recent 3D models already tentatively suggest that memory of the pre-collapse fields can be lost relatively quickly in CCSNe \citep{Varma_Mueller_Schneider:2023}, which points more towards the dynamo scenario.

There are different dynamo mechanisms that can operate in different regions of the supernova core \citep{mueller_24}. A key consideration is the degree of progenitor rotation. The cores of most supernova progenitors are not expected to rotate rapidly \citep{Heger_Woosley_Spruit:2005}, and in this case the primary candidates for dynamo field amplification processes are the turbulent small-scale dynamo, driven either by turbulent convection in the heating region \citep{Mueller_Varma:2020}, the proto-neutron star (PNS) convection zone \citep{thompson_93,Raynaud_Guilet_Janka_Gastine:2020} or shear flows at the proto-neutron star PNS surface \citep{Mueller_Varma:2020}, as well as field amplification by the standing accretion shock instability \citep{Endeve_et_al:2010,endeve_12}. Sufficiently fast progenitor rotation, induced rotation due to asymmetric downflows onto the
PNS, or the spiral mode of the 
standing accretion shock instability \citep{blondin_06} can enable further dynamo mechanisms.
This includes field amplification by an 
$\alpha$-$\Omega$ dynamo in the PNS convection zone
\citep{duncan_92,thompson_93,Raynaud_Guilet_Janka_Gastine:2020} 
or the magnetorotational instability \citep{Balbus_Hawley:1991,akiyama_03,obergaulinger_09}.

Field amplification in the context of rapid rotation has long been the primary target of magnetohydrodynamic (MHD)
supernova simulations, because strong magnetic fields can tap the reservoir of PNS rotational energy and thereby \emph{drive} the explosion. Such magnetorotational explosions
are a prime candidate for explaining ``hypernovae''
and long gamma-ray bursts \citep{Iwamoto:1998,woosley_06b} whose energies appear beyond the reach of the neutrino-driven mechanism.
Simulations of magnetorotational explosions with neutrino
transport in two \citep[e.g.,][]{Burrows_Dessart_Livne_Ott_Murphy:2007,obergaulinger_22,jardine_22} and three dimensions \citep[e.g.,][]{winteler_12,Mösta_et_al:2014,Kuroda_Arcones_Takiwaki_Kotake:2020,obergaulinger_21,Bugli_Guilet_Obergaulinger:2021,powell_23,Shibagaki:2023,powell_24} have matured considerably in recent years. However, hyperenergetic explosions only constitute a very small fraction of all core-collapse supernovae, and our incomplete understanding of rotation and magnetic fields in stellar evolution implies significant uncertainties about the requisite rapidly rotating progenitors for MHD-driven hypernovae.

In recent years, magnetic fields in CCSNe
from non-rotating or slowly rotating progenitors and with normal explosion energies have also started to receive more interest. This is clearly motivated by the quest to understand the origin of neutron star magnetic fields; but it has also been realized that magnetic fields may play a non-negligible, though subdominant dynamical role even in neutrino-driven explosions. Several 3D studies have now found indications that fields created by a small-scale dynamo in the heating region can become sufficiently strong to support the development of neutrino-driven explosions \citep{Mueller_Varma:2020,Matsumoto_et_al:2022,Nakamura_et_al:2024}. In particular, very strong pre-collapse fossil fields, which may be generated as the result of stellar mergers, could jump-start the field amplification process
and lead to early, relatively energetic neutrino-driven
explosions \citep{Varma_Mueller_Schneider:2023}. These recent simulations also suggested that relatively strong neutron star surface fields -- on the boundary between
the magnetar and pulsar regime -- may be created quite generically in supernova explosions.

The extant 3D MHD supernova simulations of non-rotating or slowly rotating progenitors still have significant limitations, however. Many simulations are still relatively short and therefore permit limited conclusions on explosion energetics and remnant properties, and a broader exploration of MHD effects in supernovae across the progenitor mass range is needed. The recent suite of 3D simulations
of progenitors between $9\msun$ and $24\msun$ by
\citet{Nakamura_et_al:2024} presents an important step in this direction, but their simulations were still limited to about $0.5\,\mathrm{s}$ of post-bounce time, and prioritized a larger number of models over grid resolution, opting for a relatively coarse resolution of $2.8^\circ$ in angle. Furthermore, despite progress in modeling the magnetic fields in the interiors of massive stars \citep{Varma_Mueller:2021, Varma_Mueller:2023},
MHD supernova simulations inevitably need to explore a larger parameter space which includes the pre-collapse field strength and configuration as additional dimensions.
Because of all these limitations, the influence of magnetic fields on the energetics of the explosion and on neutron star birth properties therefore still requires additional study.

In this paper, we therefore present a small suite of
five 3D MHD CCSN simulations to explore MHD effects across the progenitor mass range.
These progenitor models cover the range from $9.5 \msun$ to $24 \msun$ and are all non-rotating models of solar metallicity. All models are set up with the same, relatively weak pre-collapse magnetic field, and evolved with identical physics.
This approach allows us to start systematically identifying features of these CCSNe, namely trends in the properties of the explosions and the remnants. Our models complement the recent suite of \citet{Nakamura_et_al:2024} in using a higher grid resolution of $1.4^\circ$ in angle and considerably longer simulation times for exploding models.
They also use the SFHo equation of \citet{Steiner_Hempel_Fischer:2013} instead of the LS220 equation of state \citep{Lattimer_Swesty:1991} for compatibility with current neutron star mass and radius constraints.

The paper is structured as follows: In Section~\ref{sec:progenitors}, we provide an overview of our progenitor models, followed by  details on
our \textsc{CoCoNuT} MHD supernova code and
the simulation setup in Section \ref{sec:setup}. 
Sections~\ref{sec:exp_results} and \ref{sec:rem_results} provide a standard analysis
of supernova explosion and remnant properties in our simulations.
Section~\ref{sec:mag_results} focuses specifically on
the evolution of magnetic fields in the simulations.
Conclusions and implications of our findings are presented
in Section~\ref{sec:conclusion}.

\section{Progenitor Models}
\label{sec:progenitors}
Our progenitors are solar-metallicity stellar models from the set used by \citet{Mueller_Heger_Liptai_Cameron:2016}. These models are all generated with the stellar evolution code \textsc{kepler} \citep{Weaver_Zimmerman_Woosley:1978, Heger_Woosley:2010}, and do not differ in other stellar parameters such as rotation or metalicity.
Table \ref{tab:progenitors} summarizes some key progenitor properties, CCSN simulation durations and the time of bounce. Whether or not the model explodes is also indicated as this is interesting to compare with other simulations of similar progenitors \citep[e.g.][]{Burrows_et_al:2020}.

\newcolumntype{Y}{>{\centering\arraybackslash}X}
\begin{table}[htbp]
\begin{center}
    \begin{tabularx}{\columnwidth}{cYcYYc}
    \hline\hline
         Model & 
         \makecell{$M_\mathrm{ZAMS}$ \\ $(\mathrm{M}_{\odot})$} & 
         Metallicity & 
         \makecell{$t_{\mathrm{final}}$ \\ $(\textnormal{ms})$} & \makecell{$t_{\textnormal{bounce}}$ \\ $(\textnormal{ms})$} &
         \makecell{Explosion \\ $(\textnormal{Y/N})$}\\
         \hline
         s9.5 & $9.5$ & solar & 966 & 138 & Y \\
         s11.5 & $11.5$ & solar & 930 & 178 & Y\\
         s14 & $14$ & solar & 497 & 236 & N\\
         s18 & $18$ & solar & 985 & 276 & Y\\
         s24 & $24$ & solar & 1590 & 286 & Y\\
         \hline\hline
    \end{tabularx}
    \caption{Summary of progenitor models and key collapse milestones. The final simulation time, $t_{\mathrm{final}}$ is given relative to the time of bounce, $t_{\mathrm{bounce}}$, which is relative to the onset of collapse. The sum $t_{\mathrm{final}} + t_{\mathrm{bounce}}$ would yield the total simulated time. The rightmost column indicates whether the corresponding simulation produces explodes (Y), or is a failed supernova (N).}
    \label{tab:progenitors}
\end{center}
\end{table}

\begin{figure*}
    \centering
    \includegraphics[width=\textwidth]{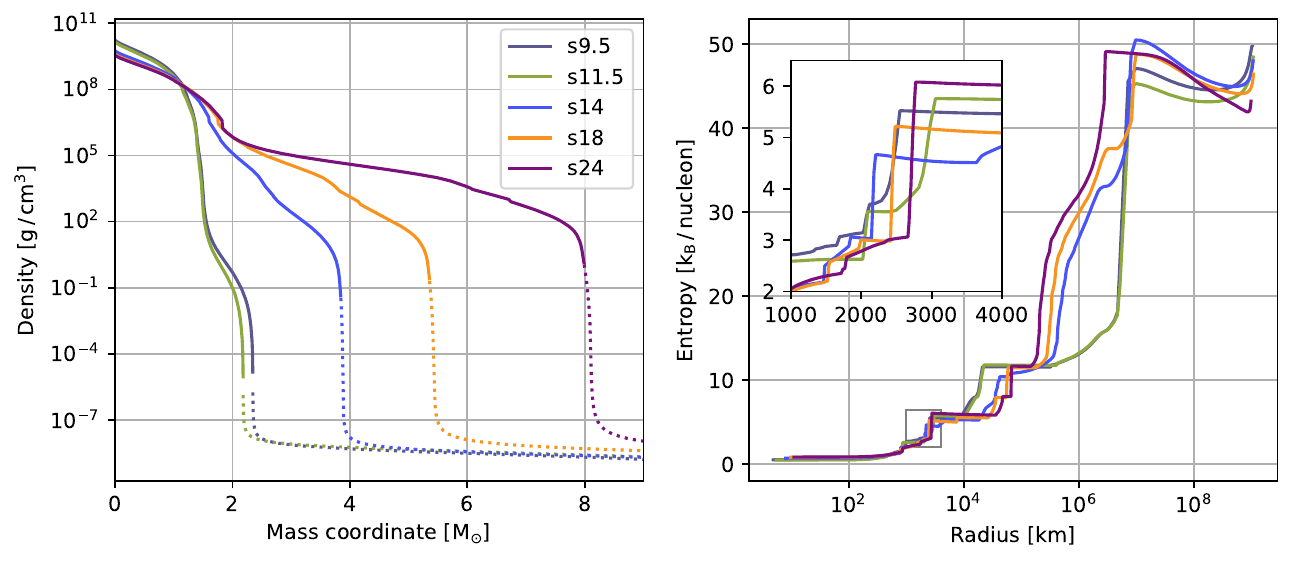}
    \caption{Density (left) and entropy (right) profiles of the progenitors used in this study (cp.\ Table \ref{tab:progenitors}). The density inside the outer hydrogen shell is shown by a solid line; a dashed line denotes the region within the hydrogen shell and is not fully shown in the plot. The inset in the right-hand figure shows, in more detail, the entropy jumps at the Si/O shell interface.}
    \label{fig:progenitor_plots}
\end{figure*}

Figure \ref{fig:progenitor_plots} shows the density structure of the progenitors produced by the \textsc{Kepler} code. Since the hydrogen shell is often of relatively low density and has a large spatial extent, the profiles are truncated at a mass coordinate of $9 \msun$. In general, the lighter models have a slightly more dense core than the heavier progenitors at the onset of collapse, but are smaller in size\footnote{We do not quantify this in terms of, e.g., the compactness parameter $\xi_{M}$ \citep{OConnor_Ott:2011} as standard choices for the fiducial mass $M$ ($1.75$ to $2.5$ \citep{OConnor_Ott:2013}) exceed the total mass on the computational grid for our two least massive progenitors.}. The sharp drop in density at the base of the H shell is similar across all progenitors

The entropy profiles show many common features, although the radius and magnitude of jumps at shell interfaces varies. Of particular interest is the region shown in the inset panel of Figure \ref{fig:progenitor_plots} which shows entropy jumps near a value of $4 \, k_\mathrm{B}$ per nucleon and corresponds to the Si/O shell interface. The infall of this shell interface is often associated with shock revival, and may be an indicator for the explodability of a star \citep{Ugliano_et_al:2012}. The position of this shell, for instance, and the corresponding free-fall time, suggests when the star will explode, while the size of the entropy jump hints at the likelihood of successful shock revival. This may explain, for instance, the failure of the s14 model to explode since it has the smallest entropy jump of the progenitors considered.

The \textsc{Kepler} stellar evolution code does not track the magnetic evolution of the stars it models. Instead, the progenitors in this work are initialized with an initial mixed poloidal/toroidal field with equal strength in both components, $B_{\mathrm{pol}} = B_{\mathrm{tor}} = 10^{8} \, \mathrm{G}$. Mathematically, this is of the form given by \citet{Varma_Mueller_Obergaulinger:2021} and similar to \citet{Obergaulinger_Just_Aloy:2018},
\begin{equation}
\begin{split}
    A^{r} & = \frac{ B_\mathrm{tor}}{r^{3}+r_{0}^{3}} r_{0}^{3}r \cos(\theta) \\
    A^{\theta} & = 0 \\
    A^{\varphi} & = \frac{B_\mathrm{pol}}{2(r^{3}+r_{0}^{3})} r_{0}^{3} r \cos(\theta)
\end{split}
\label{eqn:init_mag}
\end{equation}

\begin{figure}
    \centering
    \includegraphics[width=\linewidth]{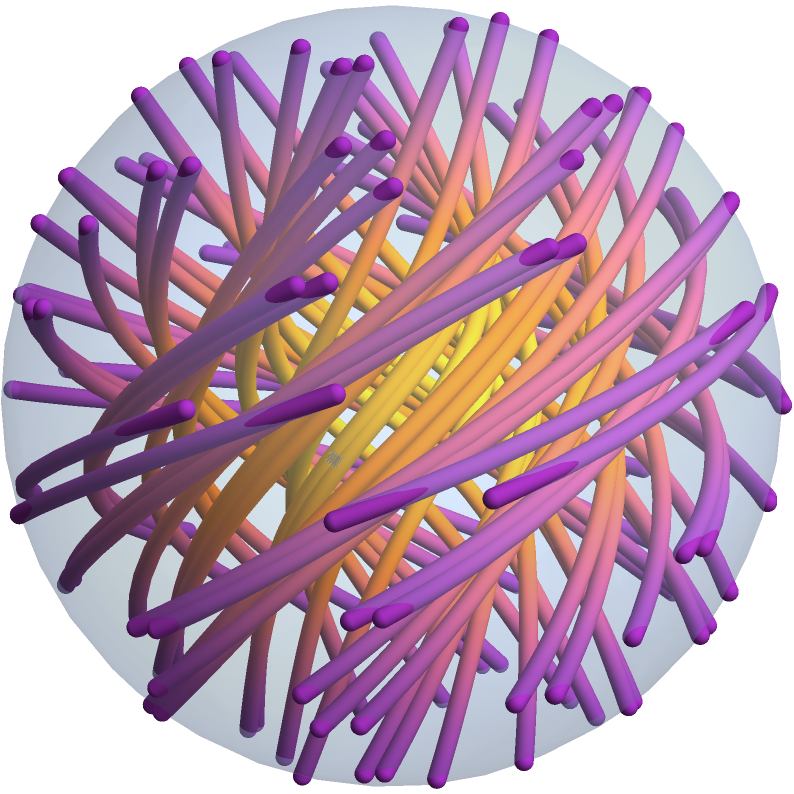}
    \caption{Field lines of the initial magnetic field for all simulations. A scale is not shown as the field geometry itself is the most pertinent feature. Field lines are rendered as tubes to make the 3D structure easier to understand and encode no physical characteristics. More yellow regions possess a stronger field strength while purple regions are weaker.}
    \label{fig:bfield_schematic}
\end{figure}

The radial scale is set to $r_{0} = 10^{3} \, \mathrm{km}$. This field exhibits significant magnetic helicity, which is expected to develop from the realistic rotation of a star; however, note that the \textsc{kepler} models do not have any intrinsic rotation, and none is inserted when the model is initialized for core-collapse. Figure \ref{fig:bfield_schematic} shows a scale-free schematic of the initial magnetic field configuration. The twisted poloidal/toroidal configuration is consistent with 3D MHD simulations of stable magnetic fields in self-gravitating plasmas \citep{Braithwaite_Spruit:2004}.

\section{Numerical Methods and Simulation setup}
\label{sec:setup}
We use the CCSN simulation code \textsc{CoCoNuT-FMT}. 
For MHD simulations, \textsc{CoCoNuT-FMT} solves the Newtonian equations of magnetohydrodynamics using finite-volume methods in spherical coordinates with a hybrid HLLC/HLLE Riemann solver for the accurate treatment of shocks. Magnetic fields are treated by adding magnetic terms to the momentum and energy equations, with additional equations for the magnetic field itself and divergence cleaning, c.p.\ \citet{Varma_Mueller:2021}. A grid resolution of $N_{r} \times N_{\theta} \times N_{\varphi} = 550 \times 128 \times 256$ is used for the radial, polar and azimuthal directions respectively. Owing to the use of spherical coordinates, the innermost $~10 \, \textnormal{km}$ are treated in 1D (i.e., spherical symmetry) to avoid an excessively short time step.

Whilst \textsc{CoCoNuT-FMT} has the capacity for general relativistic (GR) hydrodynamics, the GR variant of the MHD fluid equations is not yet implemented. Instead we use an effective gravitational potential\footnote{See \citet{Mueller_Janka_Marek:2012} for a comparison of simulations with GR, pseudo-Newtonian and Newtonian treatments of gravity.} \citep{Marek_et_al:2006, Mueller_Dimmelmeier_Mueller:2008} which has been used in numerous simulations to approximate general relativistic effects \citep{Mueller_Janka_Marek:2012, Burrows_et_al:2020}.

Neutrino transport is approximated with the fast multi-group (FMT) scheme of \citet{Mueller_Janka:2015}. The scheme provides stationary solutions to the frequency-dependent neutrino transport problem in the ray-by-ray approximation. Since it provides decent accuracy with a modest computational footprint, the FMT scheme has been employed in all recent \textsc{CoCoNuT} simulations \citep{Jakobus_et_al:2022, Mueller_Varma:2020, Powell_Mueller_Heger:2021}.

At high densities, the SFHo equation of state of \citet{Steiner_Hempel_Fischer:2013} is used. This equation of state is in decent agreement with observational and theoretical constraints \citep{Oertel_Hempel_Klahn_Typel:2017}.
At low densities, the code uses an
equation of state that includes 20 nuclear species as a perfect gas as well as electrons, positrons and photons. Nuclear reactions at low temperatures, where
nuclear statistical equilibrium does not
apply, are treated using a ``flashing''
approach \citep{rampp_02}.

\section{Explosion properties}
\label{sec:exp_results}
In this section we present key results from a standard analysis for CCSNe simulations. A summary of notable explosion and remnant properties is given in Table~\ref{tab:results}.

\begin{table*}[htbp]
\begin{center}
    \begin{tabularx}{\textwidth}{YYYYYYYY}
         \hline\hline
         Model & 
         \makecell{$E_\mathrm{expl}$ \\ $\mathrm{(10^{51} \, erg)}$} &
         \makecell{$M_\mathrm{PNS}$ \\ $\mathrm{(\mathrm{M}_{\odot})}$} &
         \makecell{$v^\mathrm{hydro}_\mathrm{kick}$ \\ $\mathrm{(km \, s^{-1})}$} & 
         \makecell{$v^\mathrm{\nu}_\mathrm{kick}$ \\ $\mathrm{(km \, s^{-1})}$} &
         \makecell{$v^\mathrm{tot}_\mathrm{kick}$ \\ $\mathrm{(km \, s^{-1})}$} &
         \makecell{$f_\mathrm{spin}$ \\ $\mathrm{(Hz)}$} &
         \makecell{$|B|_\mathrm{peak}$ \\ $\mathrm{(10^{14} \, G)}$} \\
         \hline
         s9.5 & 0.154 & 1.25 & 78 & 14 & 70 & 0.578 & 0.743 \\
         s11.5 & 0.110 & 1.25 & 75 & 15 & 87 & 4.25 & 1.85 \\
         s14 & - & 1.45 & 4.0 & 4.1 & 2.6 & 0.556 & 0.0561\\
         s18 & 0.270 & 1.64 & 258 & 54 & 253 & 15.3 & 11.6 \\
         s24 & 0.609 & 1.65 & 300 & 13 & 301 & 21.8 & 12.2 \\
         \hline\hline
    \end{tabularx}
    \caption{Summary of simulation results by the end of each simulation. The masses listed under $M_\mathrm{PNS}$ are gravitational (not baryonic) masses. The kick imparted to the PNS is decomposed into the hydrodynamical part ($v^\mathrm{hydro}_\mathrm{kick}$) and the neutrino part ($v^\mathrm{\nu}_\mathrm{kick}$);  $v^\mathrm{tot}_\mathrm{kick}$ is the vector sum of the two. Spin frequencies include both the hydrodynamic and magnetic contributions to the PNS angular momentum. The angle-averaged magnetic field is used to find the peak field strength, $|B|_\mathrm{peak}$; consequently, there may be small scale fluctuations above these peak values.}
    \label{tab:results}
\end{center}
\end{table*}

\begin{figure*}
    \centering
    \includegraphics[height=0.9\textheight]{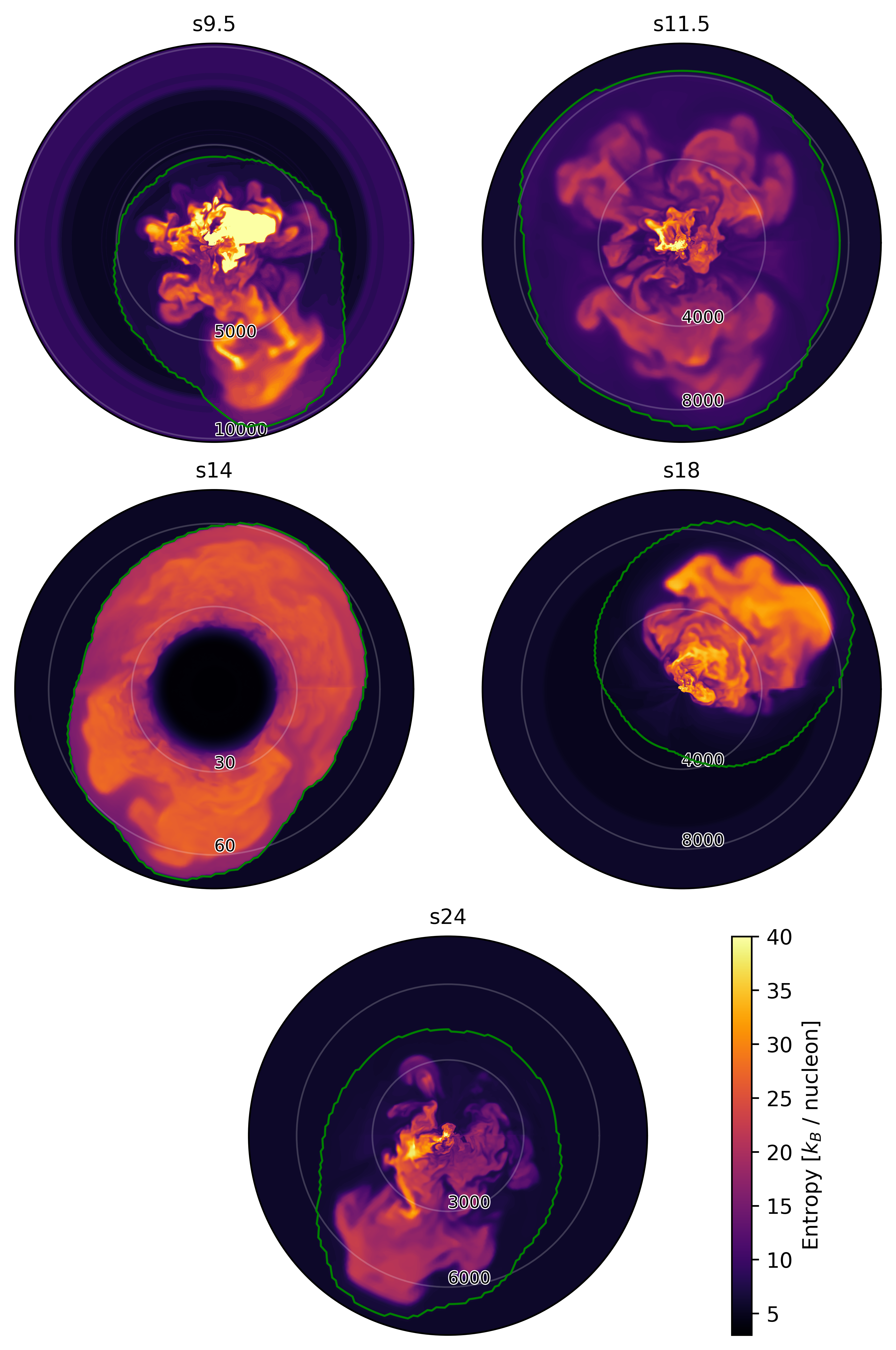}
    \caption{Slices of entropy at the end of each simulation
    through a meridional plane of arbitrary longitude ($\phi \approx 141^{\circ}$ and $\phi \approx 321^{\circ}$).
    The radius in kilometers is shown by the labeled circles; note that the radial scale is not uniform between panels. A green line denotes the position of the shock. The entropy scale is consistent between all panels, and is shown by the color bar in the lower right.}
    \label{fig:slices}
\end{figure*}

Slices of entropy at the final time step are provided in Figure \ref{fig:slices}; the corresponding times are given in Table \ref{tab:progenitors} in the $t_{\mathrm{final}}$ column. The radial scale is marked on each panel in units of kilometers and a green line shows position of the forward shock. From these slices, it is immediately evident that all models bar s14 explode; of those exploding models, most have significant asymmetry in the plane of the slice with the exception of s11.5, which is quite spherical. We contrast this with the results of \citet{Nakamura_et_al:2024} who find all sixteen of their models explode, although their use of the LS220 equation of state, which is disfavored by theoretical predictions, may explain some of these phenomenological differences (see e.g., \citet{Powell_Mueller_Heger:2021} who obtain an earlier explosion with LS220 compared to other physical EoSs).

In the s14 model, the distortion of the shock due to a spiral SASI mode \citep{Blondin_Mezzacappa_DeMarino:2003} is clearly visible in the lower half of the panel.

\subsection{Shock trajectory}
\label{subsec:shock_trajectory}
\begin{figure}
    \centering
    \includegraphics[width=\linewidth]{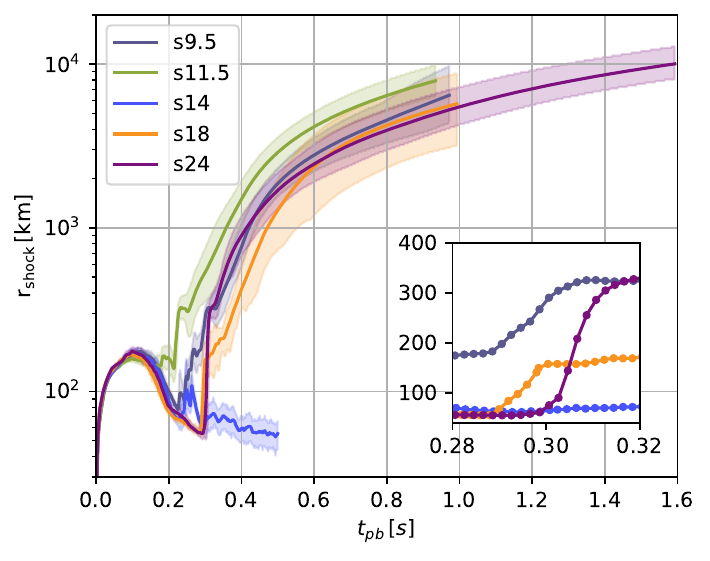}
    \caption{Shock propagation for all models. Solid lines show the mean shock radii, while the shaded regions show the central $90\%$ interval of the angle-dependent shock radius at a given time. Time is given relative to bounce. The inset panel shows a closer view of the rapid shock expansion of the s24 and s18 models, with discrete simulation outputs marked by dots.}
    \label{fig:shock_radius}
\end{figure}

Figure \ref{fig:shock_radius} presents the shock position as a function of post-bounce time for all models. Given the asymmetries which develop as the shock expands, we show the mean shock radius, as well as the central $90\%$ range across all directions. 

After bounce, the shock follows a very typical trajectory out to around \qty{\sim 200}{\kilo\meter} in all models. From here, there is some divergence in behavior with the s11.5 model showing the onset of an explosion quite early at about \qty{170}{\milli\second} post-bounce. It takes a further \qty{50}{\milli\second} for s9.5 to start exploding, at a time when the shock has receded below \qty{100}{\kilo\meter}. Another \qty{50}{\milli\second} later, the s18 and s24 models show signs of revival. The shock of the s14 model briefly expands around \qty{\sim 240}{\milli\second} post-bounce, but shock revival does not occur, with the shock slowly receding until the end of the simulation. For this model, it is anticipated that the PNS will eventually collapse to form a BH with no observable electromagnetic transient. Baring the difference in explosion time and initial radius, the shock radii of exploding models have similar trajectories.

Some models (mostly s24) appear to exhibit very sudden apparent expansion of the shock around the onset of shock revival. To some degree, this is an artifact of the shock tracking algorithm, which can get triggered by artifacts in the supersonic regions around
the infall of the Si/O shell interface. However, the issue is resolved after the accretion of the shell interface. This behavior is overaccentuated by the long simulation time scale; as the inset in Figure~\ref{fig:shock_radius} shows, shock expansion in reaction to the Si/O interface is in fact a more drawn-out process, even with the suboptimal shock tracking.

\subsection{Diagnostic explosion energy}

Following the usual definition \citep{Varma_Mueller_Schneider:2023,Mueller_Janka_Marek:2012} we compute the diagnostic explosion energy,
\begin{equation}
    E_{\textnormal{diag}} = \int\limits_{e_{\textnormal{bind}} > 0} e_{\textnormal{bind}} \,\mathrm{d}V,
\end{equation}
where the net binding energy $e_{\textnormal{bind}}$ is given by,
\begin{equation}
    e_{\textnormal{bind}} = \rho\left(\epsilon + \frac{\rho v^{2}}{2} + \Phi\right) + \frac{B^{2}}{8 \pi},
\end{equation}
in terms of the fluid internal energy $\epsilon$, velocity $v$, density $\rho$, the gravitational potential $\Phi$, and the magnetic field strength $B$.

\begin{figure}
    \centering
    \includegraphics[width=\linewidth]{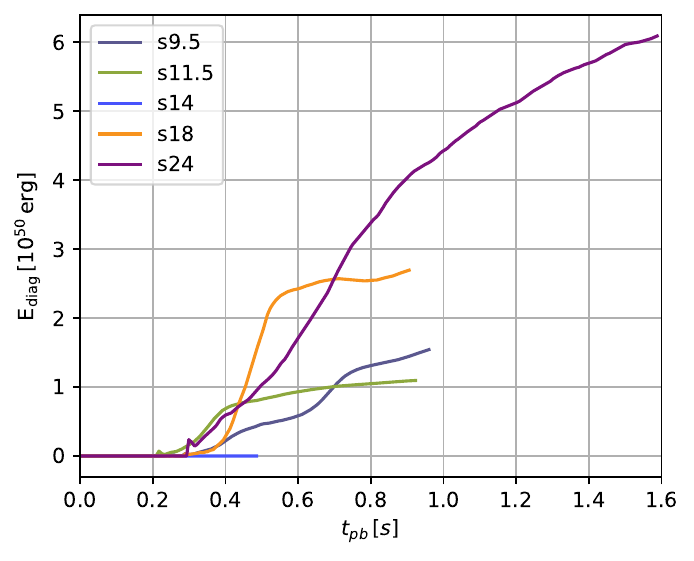}
    \caption{Diagnostic explosion energy as a function of post-bounce time for all models.}
    \label{fig:diagnostic_energy}
\end{figure}

This formal definition gives an estimate of the energy of unbound material in the supernova before it is fully determined at shock breakout. Figure \ref{fig:diagnostic_energy} shows the diagnostic explosion energy as it evolves over time in each simulation. As expected, the explosion energy starts to grow around shock revival, with all models showing rapid increases at the time their respective shocks start to expand. This is followed by a more gradual increase over hundreds of milliseconds.

Final diagnostic explosion energies (of exploding models) are in the range of $1.1 \times 10^{50} \mathrm{erg}$ to $6.1 \times 10^{50} \mathrm{erg}$. These are on the weaker side of typical CCSN explosion energies.
These energies are of the same magnitude as that obtained by \citet{Varma_Mueller_Schneider:2023} for a slightly weaker initial magnetic field.

Furthermore, none of the exploding models have yet fully converged to a final explosion energy at the end of the simulations. Our exploding models are run for about one second, although the s24 model runs longer, which is likely not long enough to arrive at a fully converged explosion energy; according to non-magnetic long-time simulations several seconds of post-bounce evolution may be required \citep{Mueller_Melson_Heger_Janka:2017, Bollig_et_al:2021,Burrows_Wang_Vartanyan:2024}. Some models appear to be approaching their asymptotic explosion energy relatively early, such as the s11.5 model, but it is difficult to predict if the flat trend of the explosion energy will continue indefinitely or if the flattening is a transient feature. The explosion energy of the long-duration s24 model steadily increases with a decreasing gradient for well over a second in a clear, well-behaved trend. It is reasonable to assume this explosion energy would likely asymptote to a final value around the canonical $10^{51} \, \mathrm{erg}$ and in decent agreement with observations.

Even models where the explosion energy seems to have flattened out could still experience growth at later times. The s9.5 model experiences multiple bursts of growth of diagnostic explosion energy, first at shock revival, and then again shortly after $600 \, \mathrm{ms}$ post-bounce. This unpredictable growth pattern makes it difficult to extrapolate for some models, highlighting the need for longer simulations in many cases. As a further example, the turnover of $E_\mathrm{diag}$ for the s18 model after about $580 \, \mathrm{ms}$ could be a long-term plateau with minor variations, however the uptick in the final $100 \, \mathrm{ms}$ of the simulation could indicate another period of increasing explosion energy. While we have simulated the s24 model for a longer time, it would be beneficial to continue all exploding models for an additional few hundred milliseconds to more firmly establish secular trends.

\subsection{Neutrino emissions}

\begin{figure}
    \centering
    \includegraphics[width=\linewidth]{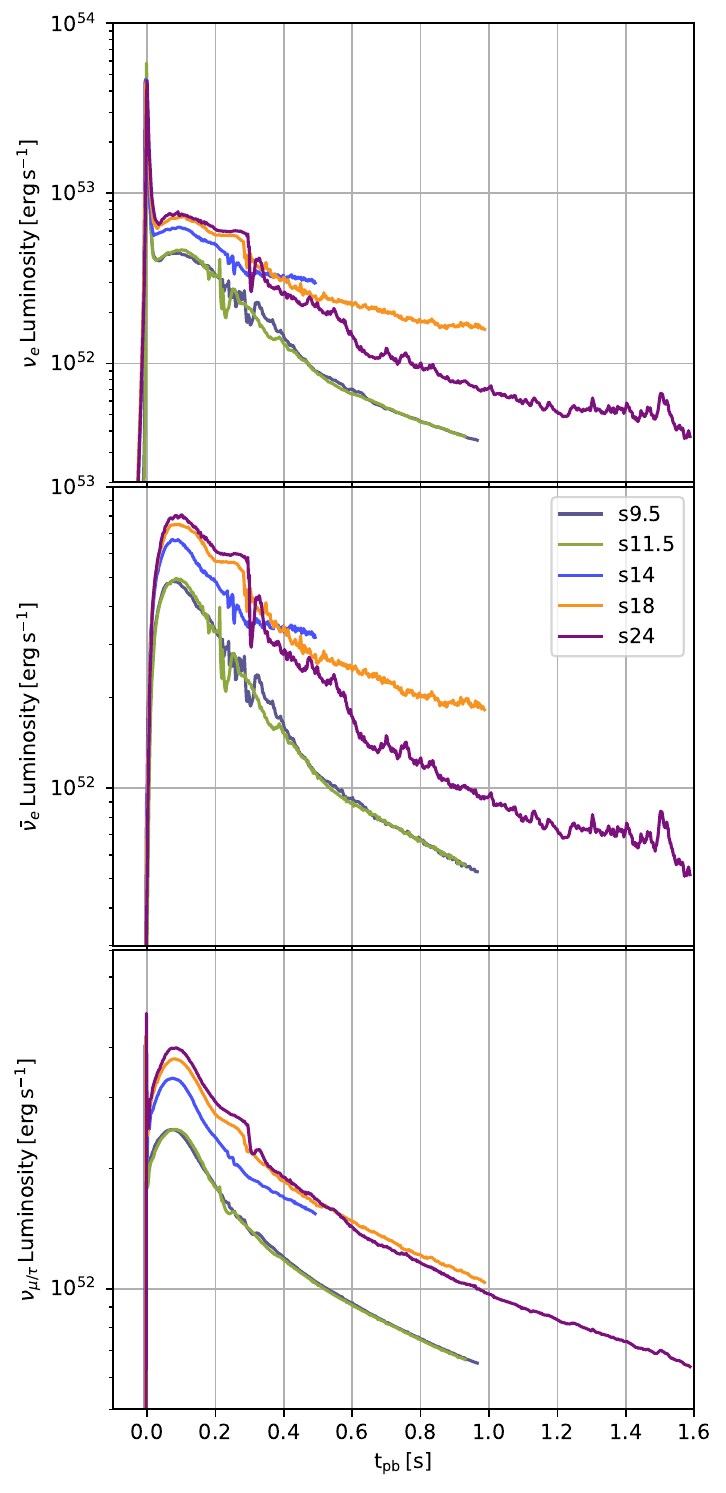}
    \caption{Neutrino luminosities for electron neutrino, electron antineutrinos and heavy-flavor neutrinos (top to bottom) as a function of post-bounce time.}
    \label{fig:neutrino_lum}
\end{figure}

Figure \ref{fig:neutrino_lum} shows neutrino luminosities of electron neutrinos ($\nu_e$), electron antineutrinos
($\bar{\nu}_e$) and heavy-flavour neutrinos ($\nu_{\mu/\tau}$). 

Normalized to the time of bounce, all models show a qualitatively similar evolution of the neutrino emission. Peak $\nu_{e}$ luminosities at bounce are close to \qty{5e53}{erg \,\second^{-1}} before dropping to a few $10^{52}\,\mathrm{erg}\,\mathrm{s}^{-1}$ in all cases. More massive progenitors maintain higher luminosities. This is in part due to the larger, hotter, and hence more luminous cores, and part due to higher accretion rates in these models (see Figure \ref{fig:mdot}). The emergence of s14 as having the strongest $\nu_{e}$ and $\bar{\nu}_{e}$ emission by the end of this simulation about $500 \, \mathrm{ms}$ post-bounce matches with a similar feature in the accretion rates.
Overall, these luminosities are consistent with typical CCSNe models.

\subsection{Accretion onto the PNS}

\begin{figure}
    \centering
    \includegraphics[width=\linewidth]{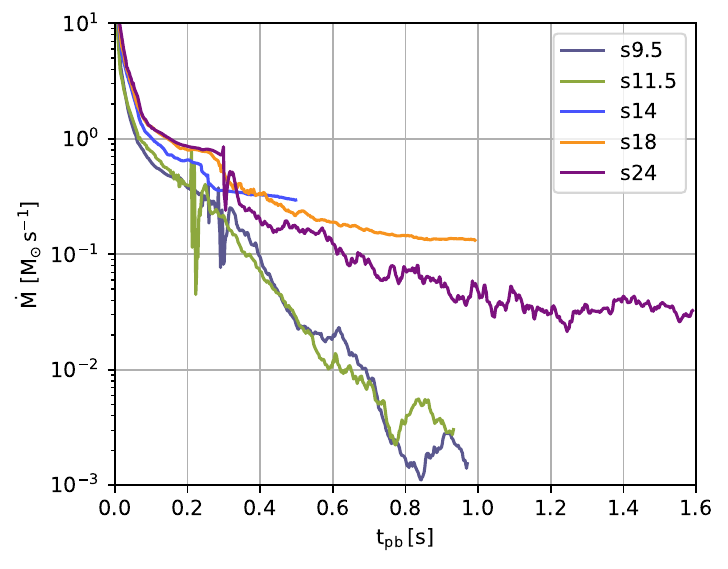}
    \caption{Mass accretion rate at a radius of $200 \, \mathrm{km}$ for all models.}
    \label{fig:mdot}
\end{figure}

Accretion rates (for inflows only) for all models are shown in Figure~\ref{fig:mdot} at a radius of $200\, \mathrm{km}$. There is a significant spread in accretion rates between models. The more massive models tend to accrete more rapidly, a fact which is also evident in the subsequent section by the gradients of the PNS mass curves in Figure~\ref{fig:ns_mass}. Of the exploding models, the s18 model accretes at the highest rate; this rate is also quite flat in the later half of the simulation. The two least massive models, s9.5 and s11.5, accrete very slowly at late times, with a rate of a few $10^{-3} \msun \, \mathrm{s}^{-1}$. This is expected 
because of the faster drop of the density outside the Si/O shell
in these progenitors.

\section{Remnant properties}
\label{sec:rem_results}
\subsection{Neutron star mass}

\begin{figure}
    \centering
    \includegraphics[width=\linewidth]{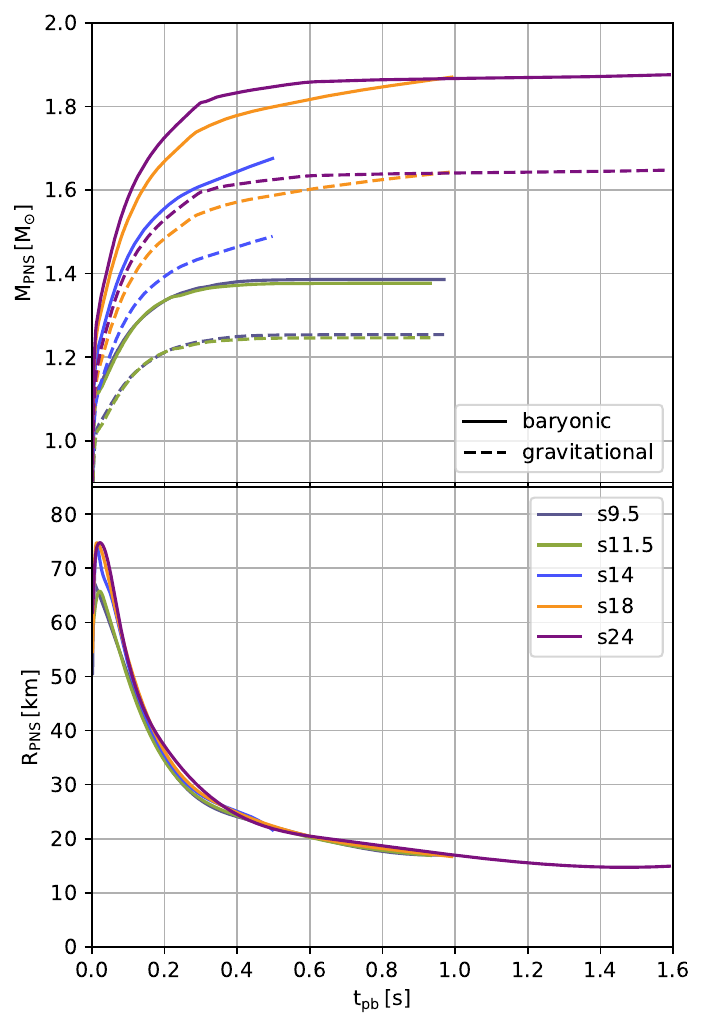}
    \caption{Top panel: PNS masses in terms of both baryonic (solid) and gravitational (dashed) masses as a function of post-bounce time for all models. Lower panel: PNS radius, also as a function of post-bounce time.}
    \label{fig:ns_mass}
\end{figure}

The evolution of the PNS mass and radius is shown in Figure \ref{fig:ns_mass}. The PNS mass quickly reaches about one solar mass at bounce, before growing rapidly at a rate that increases with progenitor ZAMS mass. The final baryonic masses vary between $1.38 \msun$ and $1.87 \msun$ with individual masses on par with other recent 3D simulations \citep{Mueller_et_al:2019,Burrows_Wang_Vartanyan:2024}. Aside from the s14 model, which shows a PNS growth rate of $0.3 \msun \, \mathrm{s}^{-1}$ at the end of the simulation, and to a lesser extent the s18 model, where the PNS mass is growing at a rate of $0.12 \msun \, \mathrm{s}^{-1}$, PNS masses are quite well converged on the timescale of the simulations.

Using the cold NS binding energy of \citet{Lattimer_Prakash:2001}, and following \citet{Mueller_et_al:2019}, we estimate the gravitational mass of the PNS with,
\begin{equation}
    M_{\mathrm{grav}} \approx M_{\mathrm{by}} - 0.084  \mathrm{M}_{\odot} \bigg(\frac{M_{\mathrm{grav}}}{\mathrm{M}_{\odot}} \bigg)^{2}.
\end{equation}
The gravitational masses range between $1.25 \msun$ and $1.64 \msun$.

The lower panel of Figure \ref{fig:ns_mass} shows the PNS radius evolution. Initially reaching between about $65 - 75 \, \mathrm{km}$ shortly after bounce, the PNS in all models shrinks in a very similar manner as time progresses.

\subsection{Remnant kick}
In 3D explosions, asphericities in the collapse and explosion may transfer a net momentum to the compact remnant.
The net momentum imparted to the remnant comes from both the hydrodynamic asymmetries \citep{Scheck_Kifonidis_Janka_Mueller:2006}, as well as asymmetries in the neutrino emission. Kicks from hydrodynamic asymmetry tend to be significantly larger than those from neutrino emission \citep{Janka_Kresse:2024}\footnote{\citet{Coleman_Burrows:2022} find the reverse ordering, with the neutrino kicks several times the corresponding hydrodynamic kick, but this may be a result of their choice of momentum flux surface \citep{Burrows_Wang_Vartanyan_Coleman:2024}.}.

Focusing first on the hydrodynamic component of the kick, the velocity imparted to the remnant neutron star is calculated by conservation of linear momentum. If the initial net momentum is assumed to be zero (which is valid since the star is stationary centered on the reference frame of the computational grid), then the net momentum outside the neutron star must cancel the momentum of the remnant entirely. This momentum, divided by the neutron star mass, gives the kick velocity, where the neutron star is defined by the region with $\rho > 10^{11} \, \mathrm{g \, cm^{-3}}$; i.e.,
\begin{equation}
    \mathbf{v}^{\mathrm{hydro}}_{\mathrm{kick}} = -\frac{1}{M_{\mathrm{PNS}}}\int_{\rho < 10^{11}\,\mathrm{g}\,\mathrm{cm}^{-1}}\mathbf{v} \rho \mathrm{dV},
\end{equation}
where $M_{\mathrm{PNS}}$ is the baryonic mass of the PNS. 
The estimation of the kick by momentum conservation has
been tested against alternative methods in the past;
see Appendix~A of \citet{Scheck_Kifonidis_Janka_Mueller:2006}.
Some previous works, for example \citet{Wongwathanarat_Janka_Muller:2013}, use the baryonic PNS mass in this calculation, while others, such as \citet{Janka_Kresse:2024}, use the gravitational mass. We have used the baryonic mass, which is arguably more accurate for estimating the asymptotic kick velocity, since later PNS cooling via isotropic neutrino emission in the PNS frame does not impact the kick velocity due to Lorentz invariance.

The contribution of asymmetric neutrino emission to the PNS kick is again calculated by means of conservation of momentum, but this time via time integration of the neutrino momentum flux at the surface, $S$, of a sphere of radius $100 \, \mathrm{km}$. Assuming forward beaming for the neutrino radiation field on this surface, these neutrino fluxes, $\mathbf{F}^{\nu}$, induce a change in velocity of
\begin{equation}
    \dot{\mathbf{v}}^{\nu}_{\mathrm{kick}} = -\frac{1}{c M_{\mathrm{PNS}}}\int_{S} \big( \mathbf{F}^{\nu_{e}} + \mathbf{F}^{\bar{\nu}_{e}} + \mathbf{F}^{\nu_{x}} \big) \mathrm{dA},
\end{equation}
which is then numerically integrated using a straightforward trapezoid method.

\begin{figure}
    \centering
    \includegraphics[width=\linewidth]{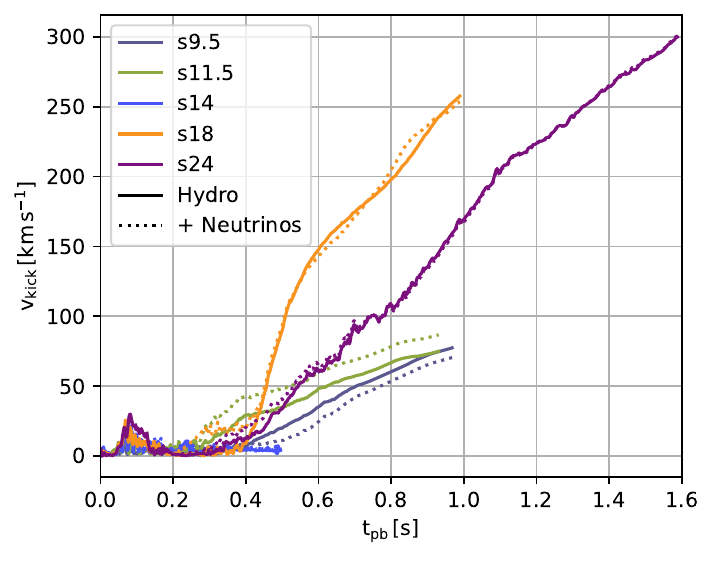}
    \caption{Magnitude of kick vector by hydrodynamic momentum conservation (solid) and with the addition of neutrino asymmetry (dotted). Final values are listed in Table \ref{tab:results}.}
    \label{fig:kick}
\end{figure}

\begin{figure*}
    \includegraphics[width=\linewidth]{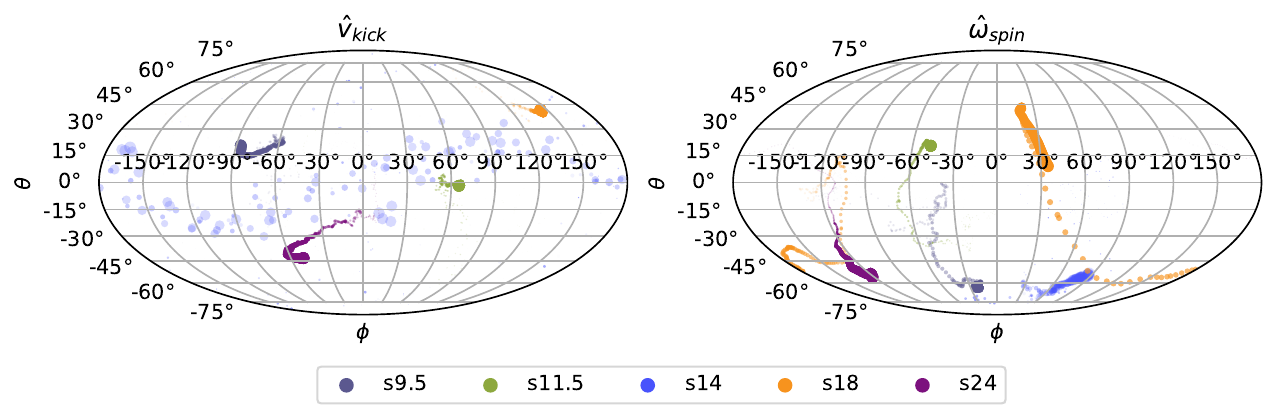}
 
    \caption{Angular trajectories of the kick (left) and spin (right) vectors. The amplitude of the vector corresponds to the transparency of the point (lighter for slower kicks/spin, darker for faster kicks/spin) while the size denotes the time, with small points growing to larger points as the simulations
    progress.}
    \label{fig:ang_spinkick}
\end{figure*}

The magnitudes of the PNS kicks are shown in Figure~\ref{fig:kick}.
We also present trajectories of the PNS kick direction 
$\hat{\boldsymbol{v}}_\mathrm{kick}$ during
the simulation in the left panel
of Figure~\ref{fig:ang_spinkick}.

The kicks are naturally small during the pre-explosion phase. 
Sinusoidal oscillations shortly after bounce correspond to the sloshing motion of the SASI found in most models. Once the explosion begins, the kick grows and its direction evolves within a relatively small solid angle; this direction is apparently random, with no correlation to the grid axis.

For the exploding models, the kick increases steadily over the simulation, with no indication of flattening out on these timescales. \citet{Burrows_Wang_Vartanyan:2024} suggest several seconds of the explosion may be required for such quantities to saturate, so the ongoing increase of our kick velocities is not surprising. The kicks are typically on the order of several tens to a few hundred kilometers per second. Most of this is due to the hydrodynamic momentum transfer and are compatible with previous results for non-magnetic 3D supernova simulations \citep[e.g.,][]{Mueller_et_al:2019,Janka_Kresse:2024, Burrows_Wang_Vartanyan_Coleman:2024, Wongwathanarat_Janka_Muller:2013}. However, we do not see large kicks (e.g. over $1000 \, \mathrm{km\,s^{-1}}$) in our limited sample. 

Our neutrino-induced kicks are subdominant and typically on the order of several tens of kilometers per second. The majority of this comes from the $\nu_{e}$ and $\bar{\nu}_{e}$ contributions, which are typically of a similar magnitude (tens of $\mathrm{km \, s^{-1}}$); the electron antineutrino provides the majority of the neutrino-induced kick in the s18 and s24 models, exceeding the contribution of electron neutrinos by a factor of $2$. Heavy-flavor neutrinos are less important for the kick, imparting only a few $\mathrm{km\,s^{-1}}$. The neutrino kicks obtained are slightly low compared to \citet{Janka_Kresse:2024} and \citet{Burrows_Wang_Vartanyan_Coleman:2024}, and much lower than \citet{Coleman_Burrows:2022}, but as mentioned previously, this may not be an accurate comparison due to the method used in
\citep{Coleman_Burrows:2022} to calculate the respective momentum contributions of matter and neutrinos. Our estimates are also a lot larger than the neutrino kicks of \citet{Wongwathanarat_Janka_Muller:2013}.

It is notable that the solid and dotted lines in Figure \ref{fig:kick} for the s18 and s24 models, representing the kick velocity without and with the effects of asymmetric neutrino emission respectively, are very similar. This is despite, for instance, neutrinos kicking the PNS in model s18 by over $60 \, \mathrm{km\, s^{-1}}$. This suggests that the net kick from neutrinos is perpendicular to the hydrodynamical kick; indeed, the angle between the neutrino and hydrodynamical kick vectors in the s18 model is about $80^{\circ}$. For the s24 model, the kicks are almost perfectly perpendicular. In the other two exploding models, the angles are $60^{\circ}$ and $140^{\circ}$ for s9.5 and s11.5, respectively. 

\citet{Janka_Kresse:2024} describe two mechanisms by which matter asymmetry can contribute to neutrino asymmetry, thereby impacting any approximate alignment of the two kicks. Firstly,  accretion downflows onto the PNS can power additional neutrino luminosity in the direction of the downflow. Secondly, dense non-accreting downflows can `block' neutrino momentum flux by absorption, with the associated momentum later returned to the PNS via the hydrodynamic mechanism; this has the effect of diminishing the neutrino kick in the direction of the dense region. These mechanisms potentially explain some alignment/antialignment of the kick vectors. Hwever our results -- and those of \citet{Janka_Kresse:2024} -- indicate a slight preference for approximately perpendicular matter and neutrino kick vectors.

\begin{figure}
    \centering
    \includegraphics[width=\linewidth]{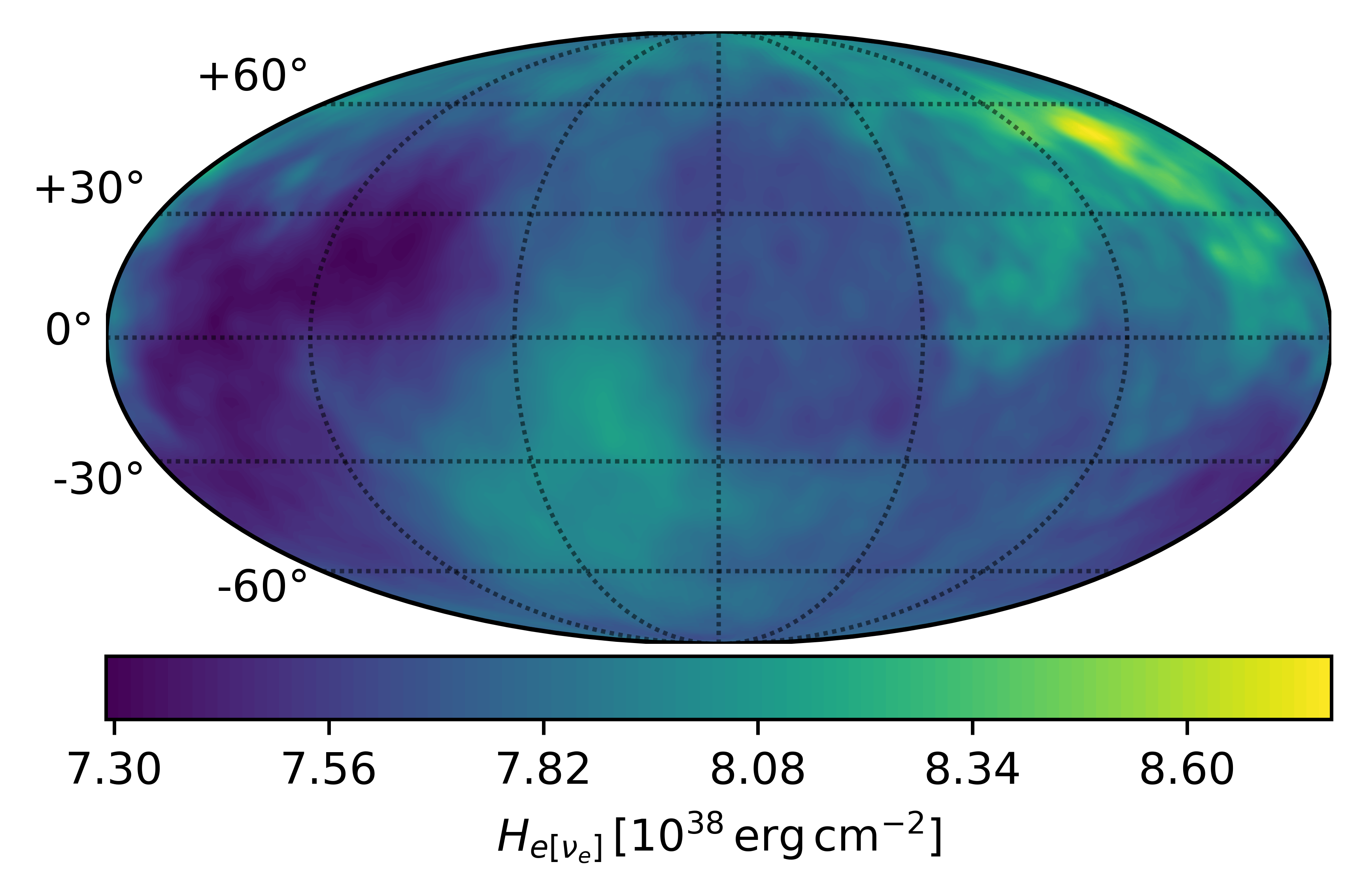}
    \caption{Asymmetric neutrino emission for the s18 model over the entire duration of the simulation. The fluence, $H_{e}$, of electron neutrinos alone is shown; i.e. no electron antineutrinos or heavy flavors. The fluence is evaluated at the PNS surface.}
    \label{fig:s18_nu_fluence}
\end{figure}

Figure~\ref{fig:s18_nu_fluence} shows the spatial distribution of electron neutrino energy fluxes integrated over the entire simulation time (i.e., the fluence) in the s18 model. There is a variation of up to $20\%$ between the minimum and maximum fluence, and the angular extent of the fluctuations are quite large -- on the order of tens of degrees latitude and longitude. Electron antineutrinos and heavy-flavor neutrinos exhibit similar distributions. 

Our PNS kick estimates match well to pulsar velocity data, e.g., as compiled recently by \citet{Willcox_et_al:2021}, although we again stress that the kicks in our simulations are not saturated, and will continue evolving for several additional seconds. Further kick evolution is unlikely to cause tensions with observations, as our kicks are slow to moderate to begin with, and so a faster kick than calculated here would still be comfortably within observational bounds.

In the case of the s14 model, which fails to explode, the final kick velocity is at most a few $\mathrm{km \, s^{-1}}$. This model will eventually produce a black hole without revival of the shock, resulting in complete collapse of the star and the formation of a black hole with the same mass as the progenitor. By momentum conservation, the remnant black hole will not have any hydrodynamic kick after the entire star has accreted. The only kick that can then be imparted is by asymmetric neutrino emission which, in our simulation, is a total of about $10 \, \mathrm{km \, s^{-1}}$ when summed over neutrino flavors. \citet{Janka_Kresse:2024} suggest kicks on the order of a few $\mathrm{km \, s^{-1}}$ for black holes, which are consistent with our results when the larger mass of the eventual black hole compared to the PNS is taken into account. The neutrino kick of the s14 model is still increasing at the end of the simulation and it is not clear how long the PNS will survive before collapse; asymmetric neutrino emission will produce a larger kick for a remnant black hole if the PNS is long-lived.

Finally, we note that our calculation of the neutrino kicks makes the assumption that the neutrino distribution function is entirely forward peaked at the evaluation radius of $100 \, \mathrm{km}$, i.e. neutrinos are acting in the free streaming limit. In a diffusive regime, this is not accurate as the off-ray momentum components are not accounted for in the correct manner. This could overstate the neutrino momentum flux. The situation may be resolved by approximately integrating over the off-ray contributions as in \citet{Mueller_Janka_Wongwathanarat:2012}, however since the effect is likely small, we elect to take the simpler upper bound for the (already subdominant) neutrino kick.

\subsection{Remnant spin}
In a similar manner to kicks, the remnant neutron star can be spun up during supernova explosions. The mechanisms behind this effect are both hydrodynamic -- simply conservation of angular momentum with decreasing radius of the body -- and magnetic, with angular momentum transferred via magnetic torques between the PNS and its surroundings.

We calculate the expected net angular momentum in the PNS by time-integration of the angular momentum flux, including the flux due to magnetic stresses, across a boundary at $\sim 50 \, \mathrm{km}$ that entirely encompasses the neutron star. 
While this sits above the typical neutron star radius of $10-20 \, \mathrm{km}$, it is likely below the final mass cut of any ongoing explosion, and consequently the extra material -- and its corresponding angular momentum -- will eventually serve to spin up/down the remnant anyway, so its inclusion is reasonable. The hydrodynamic and magnetic angular momentum fluxes, $\dot{\mathbf{J}}_{\mathrm{hydro}}$ and $\dot{\mathbf{J}}_{\mathrm{magnetic}}$, respectively, can be expressed as,
\begin{equation}
    \mathbf{\dot{J}}_{\mathrm{hydro}} = \iint(\mathbf{r} \times \mathbf{v}) \rho \mathbf{v} \cdot \mathrm{d}\mathbf{A},
\end{equation}
\begin{equation}
    \mathbf{\dot{J}}_{\mathrm{magnetic}} = \frac{1}{4 \pi}\iint(\mathbf{r} \times \mathbf{B}) \mathbf{B} \cdot \mathrm{d}\mathbf{A}.
\end{equation}

\begin{figure}
    \centering
    \includegraphics[width=\linewidth]{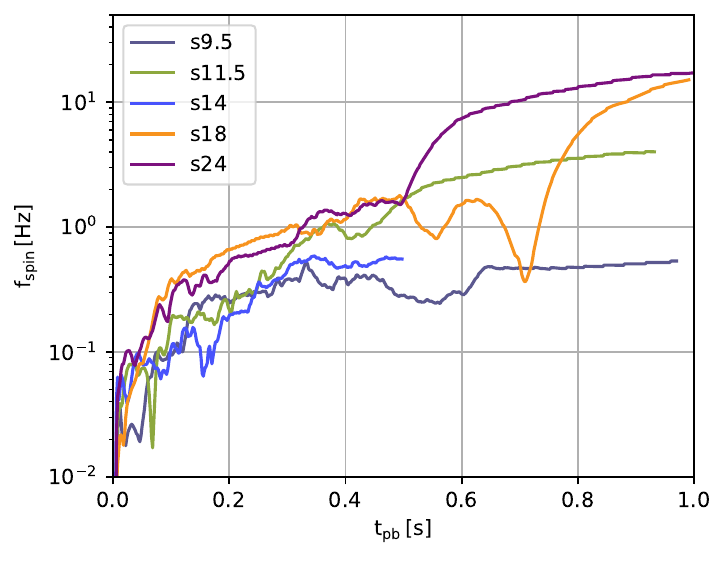}
    \caption{Spin frequency of the PNS as a function of post-bounce time for all models. Time is truncated at $1 \, \mathrm{s}$ as the longer-duration s24 model shows no substantial change after this time.}
    \label{fig:spin}
\end{figure}

We use the prescription of \citet{Lattimer_Schutz:2005} to approximate the neutron star moment of inertia, $I$, arriving at the spin frequencies shown in Figure~\ref{fig:spin} via the equation,
\begin{equation}
    f = \frac{|\mathbf{L}_{\mathrm{hydro}} + \mathbf{L}_{\mathrm{magnetic}}|}{2 \pi I}.
\end{equation}

In addition we show the trajectory of the PNS spin vector
$\hat{\boldsymbol\omega}_{\mathrm{spin}}$ in the right panel
of Figure~\ref{fig:ang_spinkick}.

With typical spin periods on the order of tens to hundreds of milliseconds, our models are representative of the observed pulsar population \citep{Igoshev_et_al:2022}. The remnants may still spin up or down due to late-time fallback \citep{Janka:2022,Mueller:2023}
and, looking forward to Section \ref{sec:mag_results}, may spin down from dipole radiation of the rotating magnetic field \citep{Igoshev_Popov_Hollerbach:2021, Mereghetti_Pons_Melatos:2015} if the surface magnetic field is strong enough -- on the order of $10^{14} - 10^{15} \, \mathrm{G}$ \citep{Thompson_Chang_Quataert:2004, Kaspi_Beloborodov:2017}.

Model s18 has a transient decrease in the spin frequency at $t=700 \, \mathrm{ms}$ post-bounce. This is consistent with a change in the spin direction, as evidenced by the $\hat{\boldsymbol\omega}_{\mathrm{spin}}$ vector in the right-hand panel of Figure~\ref{fig:ang_spinkick}, where the orange dots initially appear to traverse the lower hemisphere at latitudes below $-35^{\circ}$ before swinging up into the upper hemisphere/equatorial region rapidly (evident by the spacing on the dots during this period).

\begin{figure}
    \centering
    \includegraphics[width=\linewidth]{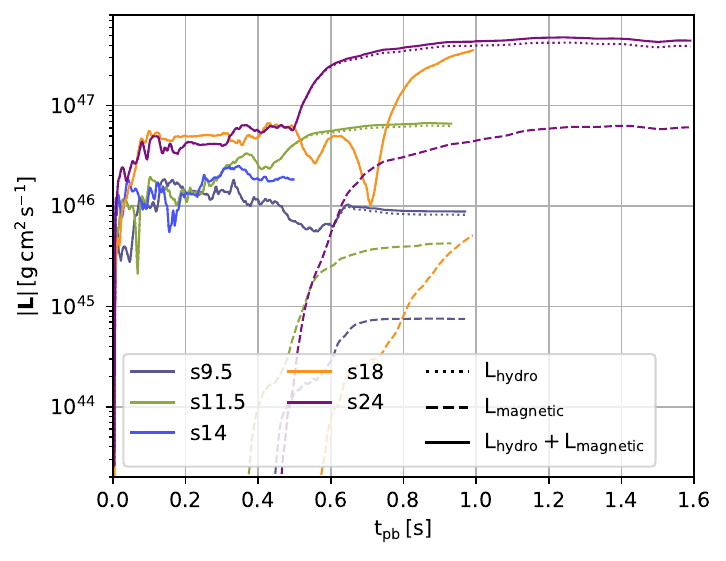}
    \caption{The angular momentum transferred to the PNS decomposed into components $L_\mathrm{hydro}$ and $L_\mathrm{magnetic}$: the hydrodynamic angular momentum flux and magnetic torques, shown as dotted and dashed lines respectively. The magnitude of the vectorial sum of both components is shown as a solid line. The contribution of magnetic torques in the s14 model is too small to be seen on this scale.}
    \label{fig:ang_mom_decomp}
\end{figure}

To differentiate the contribution of hydrodynamic advection and magnetic torques to the PNS kick, Figure~\ref{fig:ang_mom_decomp} shows the magnitude of the angular momentum gained by the PNS through each of these two channels, and how the sum of the vectors compares to the individual magnitudes.

At all times, the dominant contribution to the PNS angular momentum/spin is the hydrodynamic advection of angular momentum by accreting material. The magnetic contribution, shown by the dashed lines in Figure \ref{fig:spin}, is initially many orders of magnitude smaller than its hydrodynamic counterpart. However, once the convective dynamo in the gain region starts to effectively generate magnetic fields, the contribution of magnetic fields rises to about $10\%$. The relative contribution of magnetic torques is highest in model s24. This is enough to affect the spin of the PNS by small amount -- about $5 \, \mathrm{ms}$ on the rotation period of the s24 model. The magnetic field plays no part in the spin evolution of the non-exploding s14 model.

\begin{figure}
    \centering
    \includegraphics[width=\linewidth]{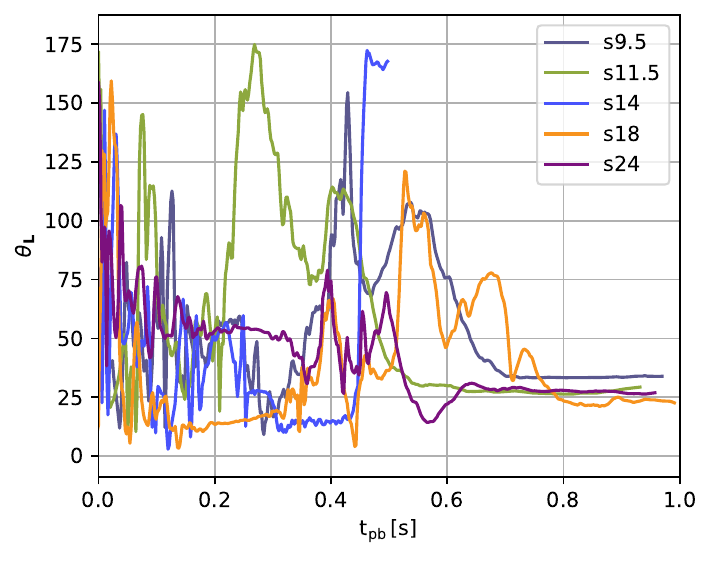}
    \caption{Angle between the net angular momentum gained by the PNS from hydrodynamic accretion and from magnetic torques, $\theta_{\mathrm{\mathbf{L}}}$, as a function of time.}
    \label{fig:lhlb_angle}
\end{figure}

Figure~\ref{fig:lhlb_angle} shows the angle $\theta_\mathbf{L}$ between the net angular momentum gained by the PNS via hydrodynamic accretion, and magnetic torques. After bounce, and for much of the simulated explosion phase, there is no discernible alignment of the two vectors. This is partly due to the stochastic nature of the accretion, which also produces the rapidly varying kicks and spins of the PNS (see Figure~\ref{fig:ang_spinkick}). The variation is also due to the relatively small size of the magnetic component, which means that even a small change in the magnetic torques can produce a large relative change in alignment. At later times when the magnetic field has grown, the alignment between these vectors stabilizes.

Interestingly, the contributions of hydrodynamic accretion and magnetic flux are almost aligned when they equilibrate near the end of the simulations ($t > 800 \, \mathrm{ms}$). Except for the s14 model, the angular momentum contributions are consistently separated only by a modest angle of $20^{\circ} - 35^{\circ}$. While near-alignment itself is not completely surprising -- as either direct settling of accreted material or magnetic torques will spin the PNS into the direction of the angular momentum of the material around the PNS -- the size of the angle and its tendency to cluster (in that no models produce better alignment with $\theta_{L}$ closer to zero) is noteworthy.

\subsection{Spin-kick alignment}
Observations of spins and kicks have revealed a statistically significant alignment between the two vectors in the pulsar population \citep{Noutsos_et_al:2012, Noutsos_et_al:2013}. This has been confirmed by ongoing surveys \citep{Yao:2021}. The mechanism behind spin-kick alignment is still not understood; it is not clear whether it results from some magnetohydrodynamic effect of the supernova engine, or from non-hydrodynamic mechanisms
(for recent overviews see \citep{Janka:2022,Mueller:2023}).
Previous hydrodynamic CCSN simulations have failed to conclusively reproduce the observed alignment \citep{Wongwathanarat_Janka_Muller:2013, Powell_Mueller_Aguilera-Dena_Langer:2023, Janka:2022, Varma_Mueller_Schneider:2023, Burrows_Wang_Vartanyan_Coleman:2024}. 

\begin{figure}
    \centering
    \includegraphics[width=\linewidth]{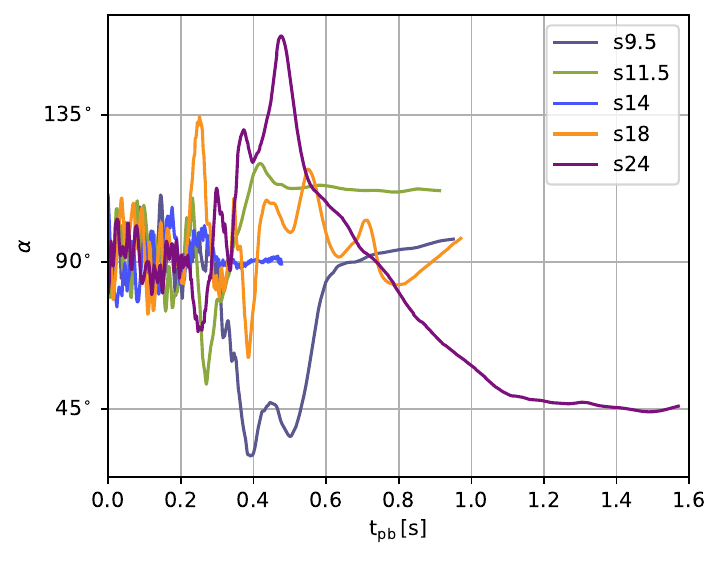}
    \caption{Angle $\alpha$ between the spin and kick vectors as a function of time. The magnitude of the vectors is not represented on this plot, only the angle between them; as such, the low amplitude oscillations soon after bounce are just stochastic noise and/or echos of SASI activity. A Savitzky–Golay smoothing filter with a window width of $40 \, \mathrm{ms}$ is applied to the data to reduce noise.}
    \label{fig:spin_kick_angle}
\end{figure}

Figure~\ref{fig:spin_kick_angle} shows the angle $\alpha$ between the spin vector and the kick vector in our simulations. At early times, it is difficult to discern any trends since the amplitude of the linear and angular momenta imparted to the neutron star are quite small -- in essence the object is being `lightly' jostled by the explosion engine in a largely stochastic manner. Once enough time has passed, non-trivial spin and kick vector magnitudes can develop and $\alpha$ exhibits clearer trends. For models s9.5 and s11.5, the preference is for misalignment of spins and kicks; the s18 model also appears to evolve towards maximum misalignment, but the angle $\alpha$ exhibits some large gradients in the final few hundred
milliseconds, which make the tendency for or against alignment ambiguous. For the s24 model, $\alpha$ evolves on a longer timescale, eventually ending up at $45^{\circ}$ -- between alignment and maximum misalignment. The SASI-dominated s14 model notably has $\alpha \approx 90^{\circ}$ for the final $200 \, \mathrm{ms}$ of the simulation despite that model not exploding.

Spin-kick alignment would require $\alpha \sim 0$. Our results
clearly do not reveal a pathway towards spin-kick alignment due to magnetohydrodynamic effects during the early explosion phase. 

\subsection{Spin-Dipole alignment}
In astrophysical NSs, the angle between the spin axis and the magnetic dipole axis, here called\footnote{This is in reference to the dipole field corresponding to $\ell=1$ in the multipole expansion.} $\theta_{1}$, impacts, among other things, the vacuum spin-down rate via dipole radiation \citep{Gunn_Ostriker:1969} and thus influences the inferred surface fields of the NS population. However, the alignment of the dipole component of the magnetic field has not yet been rigorously examined in MHD CCSN simulations. 

To determine the orientation of the dipole field, it is necessary to first extract the dipole ($\ell = 1$) components of the multipole expansion. As the magnetic field is a vector field, the expansion is done in terms of the three families of vector spherical harmonics (VSH), where we use the definitions of \citet{Barrera_et_al:1985}:
\begin{equation}
    \mathbf{Y}_{\ell m} = Y_{\ell m} \mathbf{\hat{r}}, \ \ \mathbf{\Psi}_{\ell m} = r \nabla Y_{\ell m}, \ \ \mathbf{\Phi}_{\ell m} = r \times \nabla Y_{\ell m},
\end{equation}
in terms of the scalar spherical harmonics $Y_{\ell m}$.

These admit a generic expansion of a vector-valued function in the form,
\begin{equation}
\label{eqn:vsh_decomp}
    \mathbf{B} = \sum_{\ell = 0}^{\infty} \sum_{m=-\ell}^{\ell} \bigg( C^{r}_{\ell m} \mathbf{Y}_{\ell m} + C^{(1)}_{\ell m} \mathbf{\Psi}_{\ell m} + C^{(2)}_{\ell m} \mathbf{\Phi}_{\ell m} \bigg),
\end{equation}
where any radial dependence of the field and the VSH coefficients, $C^{r}_{\ell m}$, $C^{(1)}_{\ell m}$ and $C^{(2)}_{\ell m}$, has been suppressed. These coefficients, by the usual orthogonality arguments for spherical harmonics, are,
\begin{equation}
\label{eqn:vec_ylm_coeff}
    C^{r}_{\ell m} = \int \mathbf{B} \cdot \mathbf{Y}^{*}_{\ell m} \mathrm{d}\Omega,
\end{equation}
\begin{equation}
    C^{(1)}_{\ell m} = \frac{1}{\ell(\ell + 1)} \int \mathbf{B} \cdot \mathbf{\Psi}^{*}_{\ell m} \mathrm{d}\Omega,
\end{equation}
\begin{equation}
    C^{(2)}_{\ell m} = \frac{1}{\ell(\ell + 1)} \int \mathbf{B} \cdot \mathbf{\Phi}^{*}_{\ell m} \mathrm{d}\Omega.
\end{equation}
Since we are interested in the dipole field only, we can specialize to $\ell =1$; consequently, $m=-1,0,1$. We take as granted that a dipole field has the form,
\begin{equation}
    \mathbf{B} = \frac{3 \mathbf{\hat{r}} ( \mathbf{m} \cdot \mathbf{\hat{r}} ) - \mathbf{m}}{4\pi |\mathbf{r}^{3}|},
\end{equation}
at a given radius where $\mathbf{m}$ is the magnetic dipole moment. By calculating the VSH coefficients of this exact dipole form, we find that the Cartesian components of $\mathbf{m}$ can be related
to the VSH expansion coefficients as follows,
\begin{align}
    C^{r}_{1 0} &= \frac{m_{z}}{r^{3} \sqrt{3 \pi}}  , \\
    C^{r}_{1 1} &= - \frac{m_{x} - \mathrm{i} m_{y}}{r^{3} \sqrt{6 \pi}}, \\
    C^{r}_{1 -1} &= \frac{m_{x} + \mathrm{i} m_{y}}{r^{3} \sqrt{6 \pi}}.
\end{align}
Note that only the radial $\mathbf{Y}_{\ell m}$ is needed to produce a closed system of equations; this avoids toroidal dipole components which are introduced by $\mathbf{\Psi}_{\ell m}$ and $\mathbf{\Phi}_{\ell m}$. 

The components of the dipole moment are trivially rearranged to yield a solution for $\mathbf{m}$ in terms of the VSH coefficients, which may then be calculated numerically from simulation data using Equation \eqref{eqn:vec_ylm_coeff}. For completeness, we write the resulting dipole moment vector here as,
\begin{equation}
    \mathbf{m} = r^{3}\sqrt{3 \pi}\bigg[ \frac{C^{r}_{1 -1} - C^{r}_{1 1}}{\sqrt{2}} ,  \frac{C^{r}_{1 -1} + C^{r}_{1 1}}{\mathrm{i}\sqrt{2}} , C^{r}_{1 0}\bigg].
\end{equation}

\begin{figure}
    \centering
    \includegraphics[width=\linewidth]{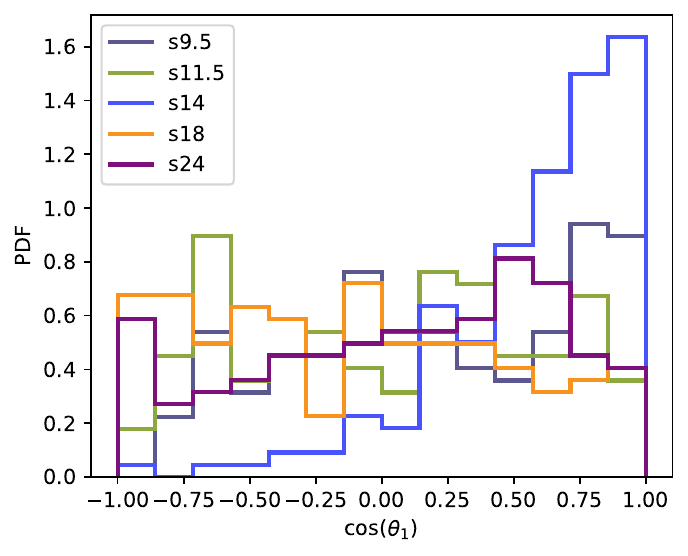}
    \caption{Probability density functions of the cosine of the angle $\theta_{1}$ between the spin and dipole moment vectors. The data are from the final $300 \, \mathrm{ms}$ post-bounce when the PNS spin axis is already quite stable in most models.}
    \label{fig:sd_hist}
\end{figure}

We find that $\theta_{1}$ is highly variable during the first second of the explosion. While the spin vector after several hundred milliseconds post-bounce has largely stabilized in the exploding models, the orientation of the magnetic dipole field fluctuates wildly on millisecond timescales. Figure \ref{fig:sd_hist} shows a histogram of the angle between the spin axis and dipole moment in the last $300 \, \mathrm{ms}$ of each simulation. There is no consistent trend in these distributions; the data for the exploding models appear nearly uniformly distributed. The variation in distributions suggests that evolution of $\theta_{1}$ is governed by the stochastic properties of the respective models, and not a unified physical process (which may emerge at a later point of the remnant's evolution).
It is interesting though that there is something of a tendency towards an alignment of the spin and dipole moment in
the non-exploding, SASI-dominated model s14.

The random variations in the direction of the magnetic dipole
moment in the exploding models indicate that the magnetic field is strongly dominated by small-scale fields with no stable global structure.
Indeed a comparison of the peak field strengths of the full magnetic field and the dipole component finds that dipole is strongly subdominant: less than $1\%$ of the total field in all models, with several as low as $0.1\%$: i.e. ranging from $10^{12} \, \mathrm{G}$ to $10^{13} \, \mathrm{G}$. The residual dipole field
is merely the result of turbulent fluctuations. Thus, different from previous
simulations \citep{Mueller:2020,Varma_Mueller_Schneider:2023}, there would seem less of a concern of overproducing neutron stars
with magnetar field strengths.

\section{Magnetic field evolution}
\label{sec:mag_results}
The analysis of the neutron star spin-up and the magnetic dipole moment in the previous section already indicated a relatively minor impact of magnetic fields on the explosion and remnant properties.
In the following, we analyze the evolution of the magnetic field further, and discuss the implications of our findings for the impact of magnetic fields on the explosion dynamics.

\subsection{Evolution of magnetic field strengths}
\label{subsec:B_energetics}

\begin{figure}
    \centering
    \includegraphics[width=\linewidth]{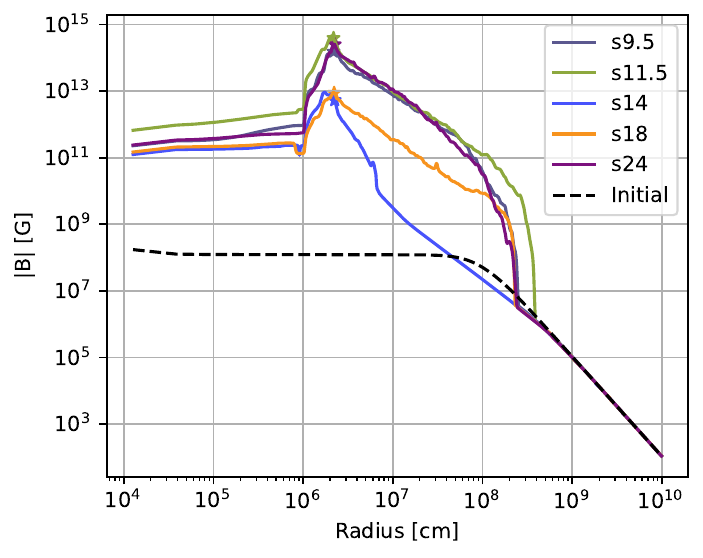}
    \caption{Radial profiles of the RMS averaged magnetic field strength at $t=500\, \mathrm{ms}$ post-bounce. The initial magnetic field strength for all models is shown as the black dashed line. A star symbol denotes the position of the PNS surface for the model of the corresponding color.}
    \label{fig:mag_radial_profiles}
\end{figure}

Radial profiles of the magnetic field strength, computed as spherical root-mean-square (RMS) averages, are given in 
Figure~\ref{fig:mag_radial_profiles} at a post-bounce time
of $500\,\mathrm{ms}$.
Each simulation is set up with an identical initial field; this is shown in Figure~\ref{fig:mag_radial_profiles} as the dashed line. Inside a radius of $500 \, \mathrm{km}$ the initial field strength is approximately constant at around $10^{8} \, \mathrm{G}$ with a power-law decay outside that region.

\begin{figure}
    \centering
    \includegraphics[width=\linewidth]{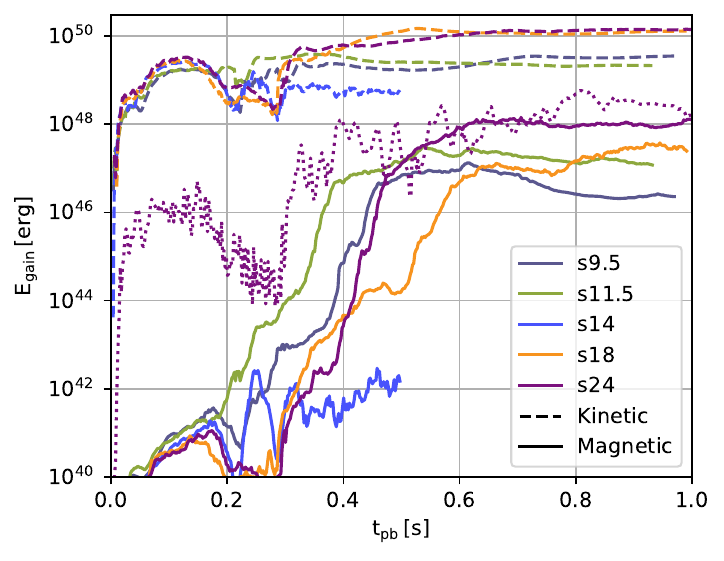}
    \caption{Time evolution of magnetic and kinetic energies in the gain region as a function of post-bounce time. Time is truncated at $1 \, \mathrm{s}$ for the s24 model as all quantities are relatively steady after this time. The dotted line shows the rotational kinetic energy in the gain region for the s24 model.}
    \label{fig:bk_energy_time}
\end{figure}

To illustrate the growth and saturation of the magnetic field we show the time evolution of the magnetic energy in the gain region in Figure~\ref{fig:bk_energy_time}. The magnetic energy $E_B$ in the gain region is simply computed as \begin{equation}
    E_B = \frac{1}{8 \pi} \int_{G}|\mathbf{B}|^{2} \mathrm{d}V.
\end{equation}

In the lead-up to bounce (not shown in Figure~\ref{fig:mag_radial_profiles}), the field increases due to compression and flux conservation. By the end of the collapse phase, it is amplified to a few $10^{11} \, \mathrm{G}$.
After bounce, the magnetic field strength increases substantially in the gain region and at the PNS surface. In a time period of $500 \, \mathrm{ms}$, the increase is by a factor in the range of $10^{3} - 10^{4}$ for the s9.5, s11.5 and s24 models. In the models that later explode,
the magnetic energies in the gain region grow by about
an order of magnitude every $100\,\mathrm{ms}$ and then
reaches maximum values of $10^{47}$-$10^{48}\,\mathrm{erg}$.
Afterwards, $E_B$ transitions to a relatively flat plateau, although the s9.5 and s11.5 models show slight decreases, while the s18 model shows a slight increase.
The s18 model experiences a somewhat more delayed growth of the magnetic field and only starts to approach magnetic saturation another $100 \, \mathrm{ms}$ later; this is why the peak field strengths are lower than the other exploding models in Figure~\ref{fig:mag_radial_profiles}. The non-exploding s14 model behaves noticeably different from the others in exhibiting much slower magnetic field amplification, with $E_B$ barely reaching above $10^{42}\,\mathrm{erg}$ by the end of the simulation.

The peak magnetic field strengths are consistently achieved at, or very near, the PNS surface, shown by a star symbol in Figure~\ref{fig:mag_radial_profiles}. The interior fields of the PNS do not grow strongly after bounce -- at most by a factor of $\sim 10$. The s14 model is less of an outlier in terms of the PNS surface field strength, which is very similar to model s18 at $500\,\mathrm{ms}$. However, it has a much weaker field inside the gain region, indicating less favorable conditions for field generation in the SASI-dominated regime, which model s14 represents.
There is no monotonic trend in the growth rate and saturation strength of the magnetic field with initial progenitor mass.

The ratio of turbulent kinetic to magnetic energy in the gain region is a key indicator for the impact of magnetic fields on the dynamics in the supernova core.
For this reason, we also show the turbulent kinetic energy $K_{\mathrm{turb}}$ in Figure~\ref{fig:bk_energy_time} for comparison to $E_B$.
The turbulent kinetic energy $K_{\mathrm{turb}}$ in the gain region is computed as,
\begin{equation}
    \label{eqn:turb_kinetic}
    K_{\mathrm{turb}} = \frac{1}{2}\int_{G}|\mathbf{v} - \mathbf{\bar{v}_{r}}|^{2} \rho \ \mathrm{d}V
\end{equation}
where $G$ is the gain region and $\mathbf{\bar{v}_{r}} = \mathbf{\bar{v}} \cdot \mathbf{\hat{r}}$ is the spherically averaged velocity in the radial direction.

The turbulent energy in the gain region rises quickly after bounce, exceeding $10^{48} \, \mathrm{erg}$ in a matter of milliseconds, before reaching somewhat steady values in the range of approximately $10^{49}$-$10^{50} \, \mathrm{erg}$.
In the non-exploding, SASI-dominated model s14, the turbulent kinetic energy in the gain region is lower and declines below $10^{49} \, \mathrm{erg}$ after the passage of the Si/O shell interface.

When $E_B$ reaches a plateau after the growth phase,
the turbulent kinetic energy still remains higher by at least two orders of magnitude at all times in our simulations for all models. This is a very similar situation as in the weak-field model of \citet{Varma_Mueller_Schneider:2023}, and different from
their strong-field case and the simulation of \citet{Mueller:2020}, where the fields reach about $40\%$ of turbulent kinetic equipartition.  In model s14, the magnetic field remains many orders of magnitude below kinetic equipartition. Clearly, the magnetic fields never reach dynamically important field strengths in our models, different from some  previous 3D MHD simulations \citep{Mueller:2020,Varma_Mueller_Schneider:2023,Matsumoto_et_al:2022}.

Why do the magnetic fields in our simulations remain further below kinetic equipartition compared to previous
models? The timing of the field amplification relative to
shock revival is likely part of the explanation.
By the time the shock starts to expand between  $200$-$300\,\mathrm{ms}$ in the exploding models
(cp.\ Figure~\ref{fig:shock_radius}), the field
strengths in the gain region are still low, and effective dynamo field amplification has just started.
Once the growth phase starts, the rate of magnetic energy increase is rapid, with most models' magnetic energy increasing by $6-7$ orders of magnitude in as little as $200$-$300 \, \mathrm{ms}$.  However, the fields are likely prevented from reaching a large fraction of equipartition because an important energy source for field generation is shut off too quickly, namely accretion power.

\begin{figure}
    \centering
    \includegraphics[width=\linewidth]{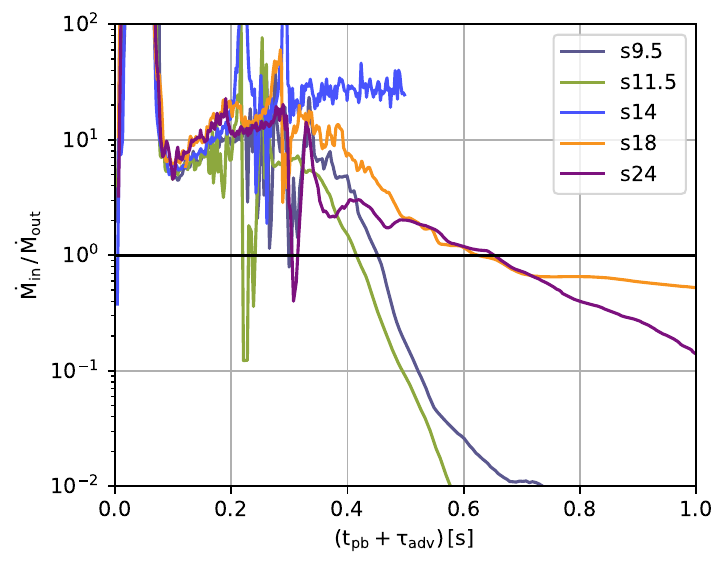}
    \caption{Ratio of the inflow and outflow rates immediately behind the shock. The horizontal axis is the post-bounce time, plus the advection timescale over the gain region.}
    \label{fig:mdot_ratio}
\end{figure}

Because the shock is revived directly after the passage of the Si/O shell interface, accretion onto the PNS starts to decline relatively quickly, thus quenching an important power source that can drive field amplification.
Qualitatively, we find that the end of the magnetic energy growth phase corresponds to an increase in the outflow from the gain region over the shock, i.e., net accretion turns into a net outflow from the PNS as the explosion develops. This is shown by Figure~\ref{fig:mdot_ratio} where the ratio of inflow and outflow rates becomes unity with the same temporal ordering as the cut-off of magnetic flux growth.
Since the inflow and outflow rates are computed behind
the shock, the advection timescale is added to the post-bounce time on the time axis to account for the traversal time between the shock surface and the primary region of magnetic flux generation near the gain radius. This advection timescale is calculated as, 
\begin{equation}
    \tau_{\mathrm{adv}} = \Delta R \sqrt{\frac{M_{\mathrm{gain}}}{2 K_{\mathrm{turb}}}},
\end{equation}
where $\Delta R$ is the width of the gain region, and the square root term is the reciprocal of a characteristic velocity based on the turbulent kinetic energy in the gain region (using the total kinetic energy produces similar results). The advection timescale is on the order of a few tens up to just over a hundred milliseconds. 

Magnetic field generation is not switched off completely when the inflow and outflow rates become equal -- after all, material is still being accreted and available for powering field amplification  -- but there is a clear signal of suppressed growth after that time. It is also significant that the two models with the sharpest cut-off of magnetic amplification and subsequent decline of magnetic field energy in the gain region, s9.5 and s11.5, have the steepest gradient of $\dot{M}_{\mathrm{in}} / \dot{M}_{\mathrm{out}}$ when the ratio equals one. By contrast, the ratio has a shallower gradient for the s18 and s24 models, which corresponds to a more drawn-out turnover of the magnetic energy. Model s18 shows continued magnetic growth after the initial sharp turnover around $0.6 \, \mathrm{s}$ post-bounce; this is consistent with the accretion rate staying relatively high ($>50\%$ of the outflow rate) for the duration of the simulation. Higher accretion rates provide a substantial energy reservoir to the dynamo and powers further magnetic growth in this simulation. Finally, the slow turnover of the s24 model over $>100 \, \mathrm{ms}$ starting at about $470 \, \mathrm{ms}$ post-bounce may be explained by the accretion ratio being small earlier in this model compared to others, i.e., before the ratio reaches unity. The ratio drops as low as $2$ roughly when the gradient of magnetic energy starts decreasing, although it is significantly lower than the comparable s18 model for a significant duration beforehand.

The non-exploding s14 model is evidently not subject to this condition as the internal dynamics of the gain region are quite different. Inhibition of growth in this model may be a similar phenomenon to the inhibition of convection in the case of fast accretion through the gain region \citep{foglizzo_06}. As pointed out by \citet{foglizzo_06}, advection can stabilize a flow against convection if the ratio between the advection time scale and the buoyancy time scale is sufficiently small. For the small-scale dynamo a similar phenomenon appears plausible.

Although the magnetic field energy in the gain region stagnates when the accretion rate drops, this is still to be understood not as a cut-off dynamo amplification altogether, but rather as saturation at a low level, set by some form of balance between continuing amplification processes (see Section~\ref{sec:spatial_analysis}) and turbulent reconnection. Due to the changed character of the flow during the explosion phase, saturation may simply happen at a rather low fraction of kinetic equipartition. Our simulation data are consistent with the possibility that different energy scales may determine the magnetic field saturation level during the explosion phase.
Since the curves of turbulent kinetic energy and magnetic field energy in the gain region all reach something of a plateau separated by 2-3 orders of magnitude, it may very well be the kinetic energy in the gain region that determines the saturation field strength, but with a very small proportionality constant. Alternatively, if magnetic field amplification is primarily driven by non-radial shear flows due to the emerging (differential) rotation at the PNS surface and its vicinity, the relevant energy scale would be the rotational energy, leading to saturation field strengths of order \citep{Burrows_Dessart_Livne_Ott_Murphy:2007},
\begin{equation}
\label{eqn:B_sat_approx}
    B \sim \sqrt{4 \pi \epsilon \rho r^{2} \omega^{2}},
\end{equation}
where the non-dimensional constant $\epsilon$ is expected to be somewhat smaller than unity (e.g., $\epsilon \sim 0.1$).

For the s24 model (solely, to avoid clutter), we therefore also compare $E_B$ to the rotational energy $E_\mathrm{rot}$ in the gain region,
\begin{equation}
     E_\mathrm{rot} = \frac{1}{3} \rho r^{2} \omega^{2}.
\end{equation}
Here the shell-averaged angular velocity $\omega$ is calculated from the shell-averaged angular momentum
as $ \omega = \left|\int \mathbf{r}\times \rho \mathbf{v}\,\mathrm{d}\Omega\right|/I$, where $I$ is the moment of inertia of the shell (assuming an isotropic density distribution). 
The magnetic and rotational energies are of comparable scale for this model. Our models are thus consistent with saturation magnetic field strengths during the explosion phase being either set as a small fraction of kinetic equipartition or a relatively high fraction of rotational equipartition.

\subsection{Spatial analysis of field amplification}
\label{sec:spatial_analysis}
For better understanding the nature of field amplification in the explosion phase and highlighting the importance of accretion down to the PNS surface, it is useful to perform a spatial analysis of field amplification.
The local growth of the magnetic field is described by the induction equation,
\begin{equation}
    \label{eqn:induction}
    \frac{\partial \mathbf{B}}{\partial t} = \nabla \times (\mathbf{v} \times \mathbf{B}) + \eta \nabla^{2} \mathbf{B},
\end{equation}
where $\eta$ is the magnetic diffusivity, implicitly taken  to also include the effects of numerical reconnection in the turbulent environment of the PNS. 
Here we only focus on the amplification term in the induction equation. From the induction equation, we can obtain the volumetric rate of magnetic energy generation, $\dot{e}_B=\dot{\mathbf{B}}\cdot\mathbf{B}/(4\pi) $.

\begin{figure}
    \centering
    \includegraphics[width=\linewidth]{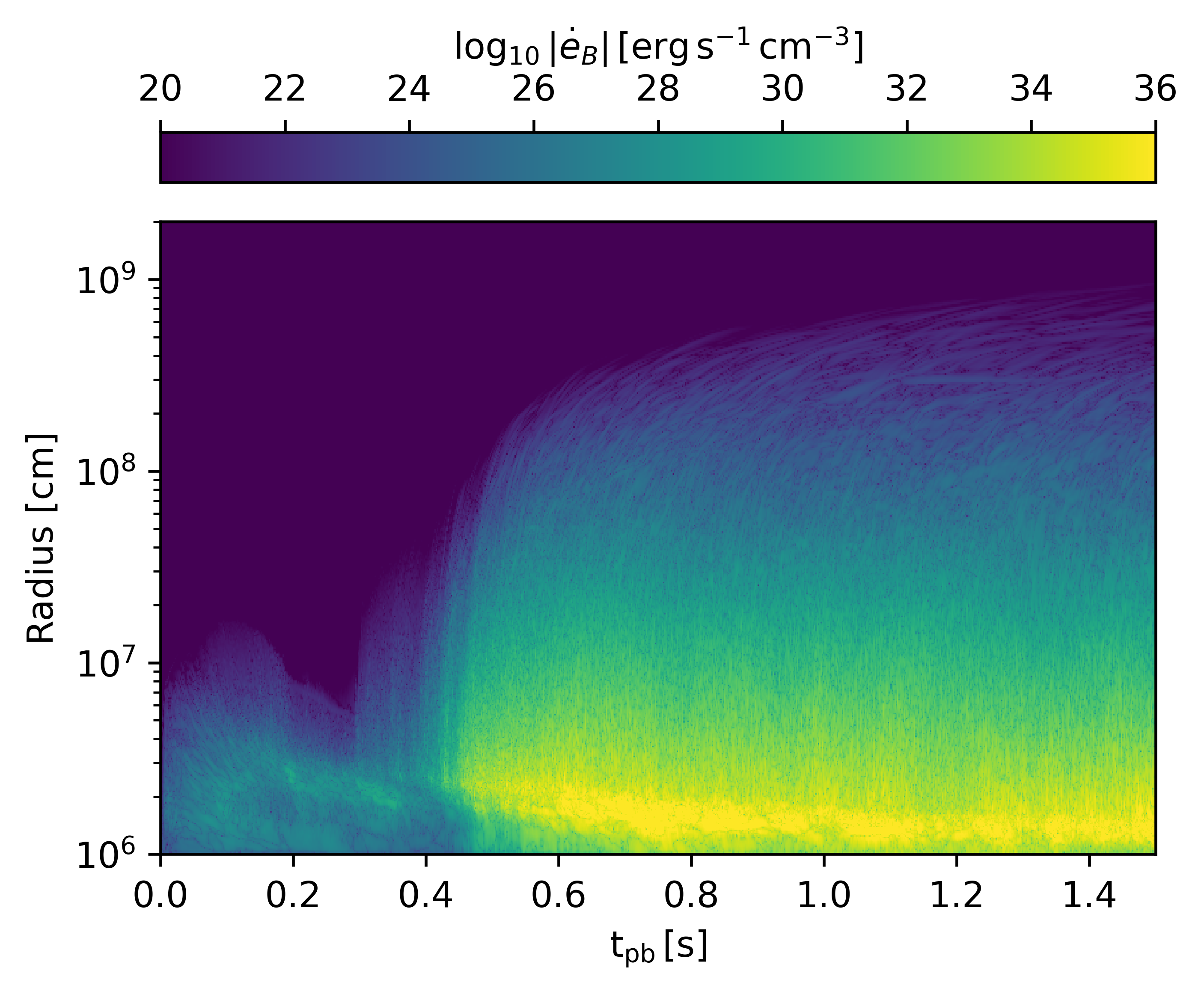}
    \caption{Magnitude of the spherically-averaged volumetric magnetic energy generation rate as a function of time and radius for the s24 model.}
    \label{fig:mag_rate}
\end{figure}

Figure~\ref{fig:mag_rate} shows the absolute value of the spherically averaged energy generation rate $\dot{e}_B$ as a function of radius and time for model s24.
The largest values of $|\dot{e}_B|$ are clearly encountered
at the base of the gain region. It must be noted that
$\dot{e}_B$ can be strongly positive or negative in this region, with changes on small spatial and temporal scales
(giving rise to spotty patterns in Figure~\ref{fig:mag_rate}),
indicating that there is also feedback of the magnetic fields onto the flow that expends their energy. Nonetheless, Figure~\ref{fig:mag_rate} clearly identifies the PNS surface region as the most relevant for field amplification.

\begin{figure}
    \centering
    \includegraphics[width=\linewidth]{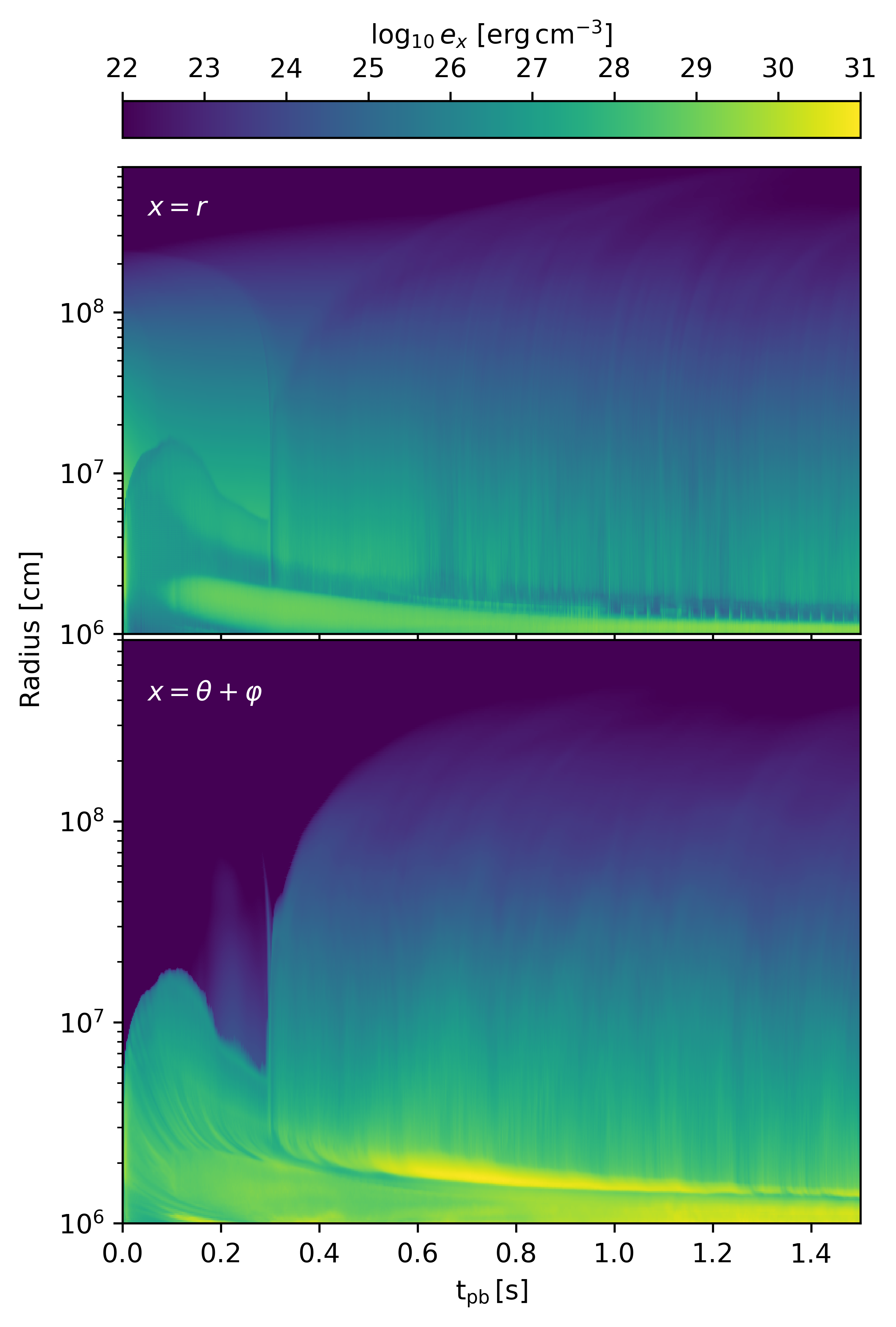}
    \caption{Kinetic energy densities for radial velocity (top panel) and angular velocities (bottom panel) for the s24 model.}
    \label{fig:vel_power}
\end{figure}

This clear localization of field amplification processes implies that the primary driver of magnetic fields are shear flows in the PNS surface region rather than turbulence in the bulk of the ejecta region. We illustrate the connection to shear flows by showing the spherically averaged radial and non-radial kinetic energy of the flow 
$e_r$ and $e_{\theta+\varphi}$ in the fluid,
\begin{equation}
e_r=\frac{1}{2}\int \rho v_r^2\,\mathrm{d}V
\end{equation}
\begin{equation}
e_{\theta+\varphi}=\frac{1}{2}\int \rho (v_\theta^2+v_\varphi^2)\,\mathrm{d}V
\end{equation}
The components $e_r$ and $e_{\theta+\varphi}$
are shown in Figure~\ref{fig:vel_power}, also for model s24. 
The shock trajectory can be seen as a faint color discontinuity in the upper panel for $e_r$, and somewhat more clearly in the lower panel for $e_{\theta+\varphi}$.
The top panel shows, with darker shades, lower radial kinetic energy at the bottom of the gain region and in the convectively stable PNS surface region, which contracts to a radius below $20\,\mathrm{km}$ over the course of the simulation.
Below this, there is the PNS convection zone shown in lighter shades.

The lower panel also shows a very thin zone at the PNS surface with small $e_{\theta+\varphi}$, but in contrast to the radial kinetic energy, the transverse kinetic energy is large immediately above it, at the bottom of the gain region. A band of high transverse kinetic energy is clearly visible in yellow at times later than about $500 \, \mathrm{ms}$ post-bounce. The region of large kinetic energy in transverse flows corresponds to the region of strong magnetic energy generation in Figure~\ref{fig:mag_rate}. This clearly points to shear flows at the bottom of the gain region -- and hence to the important of accretion -- for field amplification process, rather than to the turbulence in the bulk of the ejecta region or differential rotation in the PNS.

\section{Conclusions}
\label{sec:conclusion}
In this work we have presented a set of five long-time 3D MHD CCSN simulations for non-rotating progenitors with ZAMS masses
$9.5 \msun$, $11.5 \msun$, $14 \msun$, $18 \msun$ and $24 \msun$. Except for the $14 \msun$ model, all of these successfully explode and have been evolved to about $1\,\mathrm{s}$ after bounce or more. This dimorphism even among progenitors close in ZAMS mass is consistent with other recent systematic studies (e.g., \citet{Kresse_Ertl_Janka:2021, Vartanyan_Burrows:2023}). The non-explosion of a relatively low-mass progenitor mirrors the trends in \citet{Vartanyan_Burrows:2023,Burrows_Wang_Vartanyan:2024}.

The first 3D MHD CCSN simulations of non-rotating progenitor had indicated that magnetic fields may somewhat aid the development of explosions, but had left open the question whether magnetic fields significantly boost explosion energies \citep{Varma_Mueller_Schneider:2023}.
For the exploding models we find final diagnostic explosion energies between $1.1 \times 10^{50} \, \mathrm{erg}$ and $6.1 \times 10^{50} \, \mathrm{erg}$, but in most cases the explosion energies are still increasing at the end of the simulation. This is consistent with the findings of \citet{Mueller:2015,Mueller_Melson_Heger_Janka:2017,Bollig_et_al:2021} and \citet{Burrows_Wang_Vartanyan:2024} that several seconds of simulated time post-bounce may be required to achieve full saturation of explosion properties. Our simulations thus suggest that 3D MHD supernova simulations do \emph{not} produce different explosion energies compared to previous non-magnetic 3D CCSN simulations of non-rotating progenitors.

The explosions impart PNS kicks on the order of $\mathcal{O}(100 \, \mathrm{km \ s^{-1}})$ with subdominant neutrino components, and angular momentum corresponding to birth spin periods of $55 \, \mathrm{ms}$ to $1.8 \, \mathrm{s}$. The evolution of the kicks and spins are comparable to previous non-magnetic supernova simulations. Our simulations also do not show any preference for spin-kick alignment. So far, magnetic fields do not appear to provide a mechanism for establishing the observed spin-kick alignment \citep{Noutsos_et_al:2012, Noutsos_et_al:2013} during the early phase of the explosion. As in previous supernova simulations \citep{Varma_Mueller:2023, Burrows_Wang_Vartanyan_Coleman:2024}, the mechanism for the alignment remains elusive. Magnetic stresses play a non-negligible, but subdominant role in spinning up the PNS compared to hydrodynamics angular momentum fluxes.

The RMS magnetic fields at the PNS surface consistently reach the $10^{14} - 10^{15} \, \mathrm{G}$ range for exploding models. No trends in the surface fields with progenitor mass can be identified.
The dipole field strengths are considerably smaller and fall in the range $10^{12} - 10^{13} \, \mathrm{G}$. This is compatible with the birth dipole fields of young pulsars \citep{Kaspi_Beloborodov:2017}. However, it must be borne in mind that the surface fields during these early epochs may still evolve considerably, e.g., due to processes like field burial \citep{Romani:1990}.

Our simulations present a more tempered picture about the dynamical role of magnetic field in supernovae of non-rotating progenitors than the first 3D simulations of magnetically-aided neutrino-driven explosions may have suggested \citep{Mueller:2020,Matsumoto_et_al:2022,Varma_Mueller_Schneider:2023}. Significantly, we find saturation of the magnetic energy in the gain region of at most $1\%$ of turbulent kinetic energy. For the non-exploding $14 \msun$ model, which shows
sustained SASI activity, field amplification inside the gain region is even less efficient, though the field strengths at the PNS surface are comparable to the other models.

A key difference to the case considered by \citet{Mueller:2020} is that the current 3D models experience shock revival directly after the infall of the
Si/O shell interface, whereas the explosion was delayed 
further in \citet{Mueller:2020}. As a result, our models are missing the critical phase of favorable growth conditions in the pre-explosion phase that allowed the field strength to reach a large fraction of equipartition.
A more modest dynamical role of magnetic fields is also consistent with the recent results of \citet{Nakamura_et_al:2024} who found only a small impact of MHD effects on explosion energies. Importantly, our considerably longer simulations show that magnetic fields remain relatively unimportant on longer time scales in the explosion phase. With subsiding accretion, further amplification by shear flows at the PNS surface -- the main driver of late-time field amplification in our simulations -- seems not be effective and the magnetic energy in the gain region stagnates at $10^{48}\,\mathrm{erg}$ or less. For supernova modeling this is an important, if unspectacular finding. For many purposes, e.g., the investigation of explosion energies and nucleosynthesis, neglecting magnetic fields in supernova simulations of non-rotating progenitors may remain a justifiable approximation.

Nevertheless, further investigation of MHD effects in supernova of non-rotating progenitors remains important. Our set of long-term MHD simulations is still relatively small compared, e.g., to the recent sets of non-magnetic 3D simulations of \citet{Burrows_Wang_Vartanyan:2024} or the much shorter, lower-resolution MHD study of \citet{Nakamura_et_al:2024}. Cases with a more significant role of magnetic fields 
around shock revival as in the model \citet{Mueller_Varma:2020} should also be investigated in long-time simulations of the explosion phase.

Importantly, the same caveats about resolution and non-ideal MHD effects hold that were already outlined in \citet{Mueller_Varma:2020}: Do the simulations correctly capture the high-Reynolds number limit in the relevant regime of magnetic Prandtl number, and could field amplification be more efficient at high resolution as in the recent MHD convection simulations of \citet{Leidi_et_al:2023}? Furthermore, MHD supernova simulations still offer no explanation for the dichotomy in neutron star birth magnetic fields between pulsars and magnetars \citep{Kaspi_Beloborodov:2017}. Longer, more systematic MHD supernova simulations, resolution studies and code comparisons remain essential for comprehensively understanding the role of magnetic fields in normal, neutrino-driven core-collapse supernovae.

\begin{acknowledgments}
This research is supported by an Australian Government Research Training Program (RTP) Scholarship.
BM acknowledges support by
the Australian Research Council (ARC)
through grants FT160100035 and DP240101786.
We acknowledge computer time allocations from Astronomy Australia Limited's ASTAC scheme, the National Computational Merit Allocation Scheme (NCMAS), and
from an Australasian Leadership Computing Grant.
Some of this work was performed on the Gadi supercomputer with the assistance of resources and services from the National Computational Infrastructure (NCI), which is supported by the Australian Government, and through support by an Australasian Leadership Computing Grant.  Some of this work was performed on the OzSTAR national facility at Swinburne University of Technology.  OzSTAR is funded by Swinburne University of Technology and the National Collaborative Research Infrastructure Strategy (NCRIS). 
\end{acknowledgments}

\appendix

\bibliographystyle{apsrev4-2}
\bibliography{bibliography.bib}

\end{document}